\documentclass[
 reprint,
 superscriptaddress,
 longbibliography,
 nofootinbib,
 amsmath,
 amssymb,
 aps,
 prd,
 floatfix,
 showkeys
]{revtex4-2}

\usepackage{amsmath, amssymb}
\usepackage{graphicx}
\usepackage{hyperref}
\usepackage{times}
\usepackage{xcolor}
\usepackage{bm}
\usepackage{multirow}
\usepackage{orcidlink}

\definecolor{arsenii}{RGB}{139,0,139}
\definecolor{andrea}{RGB}{0,153,76}
\definecolor{dmitry}{RGB}{12,127,172}
\definecolor{comments}{RGB}{255,111,0}
\definecolor{attention}{RGB}{255,0,0}

\begin{document}

\title{Simulation-based inference for Precision Neutrino Physics through Neural Monte Carlo tuning}

\newcommand{\affiliationJINR}{Joint Institute for Nuclear Research, Dubna, Russia}
\newcommand{\affiliationINR}{Institute for Nuclear Research of the Russian Academy of Sciences, Moscow, Russia}

\author{A. Gavrikov${}^{\orcidlink{0000-0002-6741-5409}}$}
\email{arsenii.gavrikov@pd.infn.it}
\affiliation{INFN, Sezione di Padova e Università di Padova, Dipartimento di Fisica e Astronomia, Italy}

\author{A. Serafini${}^{\orcidlink{0000-0001-9191-661X}}$}
\email{andrea.serafini@pd.infn.it}
\affiliation{INFN, Sezione di Padova e Università di Padova, Dipartimento di Fisica e Astronomia, Italy}

\author{D. Dolzhikov${}^{\orcidlink{0000-0002-5070-7501}}$}
\email{ddolzhikov@jinr.ru}
\affiliation{\affiliationJINR}
\affiliation{\affiliationINR}

\author{A. Garfagnini${}^{\orcidlink{0000-0003-0658-1830}}$}
\affiliation{INFN, Sezione di Padova e Università di Padova, Dipartimento di Fisica e Astronomia, Italy}

\author{M. Gonchar${}^{\orcidlink{0000-0002-6820-9471}}$}
\affiliation{\affiliationJINR}
\affiliation{\affiliationINR}

\author{M. Grassi${}^{\orcidlink{0000-0003-2422-6736}}$}
\affiliation{INFN, Sezione di Padova e Università di Padova, Dipartimento di Fisica e Astronomia, Italy}

\author{R. Brugnera}
\affiliation{INFN, Sezione di Padova e Università di Padova, Dipartimento di Fisica e Astronomia, Italy}

\author{V. Cerrone}
\affiliation{INFN, Sezione di Padova e Università di Padova, Dipartimento di Fisica e Astronomia, Italy}

\author{L. V. D'Auria}
\affiliation{INFN, Sezione di Padova e Università di Padova, Dipartimento di Fisica e Astronomia, Italy}

\author{R. M. Guizzetti} 
\affiliation{INFN, Sezione di Padova e Università di Padova, Dipartimento di Fisica e Astronomia, Italy}

\author{L. Lastrucci}
\affiliation{INFN, Sezione di Padova e Università di Padova, Dipartimento di Fisica e Astronomia, Italy}

\author{G. Andronico}
\affiliation{INFN, Sezione di Catania e  Università di Catania, Dipartimento di Fisica e Astronomia, Italy}

\author{V. Antonelli}
\affiliation{INFN, Sezione di Milano e Università degli Studi di Milano, Dipartimento di Fisica, Italy}

\author{A. Barresi}
\affiliation{INFN, Sezione di Milano Bicocca e Dipartimento di Fisica Università di Milano Bicocca, Italy}

\author{D. Basilico}
\affiliation{INFN, Sezione di Milano e Università degli Studi di Milano, Dipartimento di Fisica, Italy}

\author{M. Beretta}
\affiliation{INFN, Sezione di Milano e Università degli Studi di Milano, Dipartimento di Fisica, Italy}

\author{A. Bergnoli}
\affiliation{INFN, Sezione di Padova e Università di Padova, Dipartimento di Fisica e Astronomia, Italy}

\author{M. Borghesi}
\affiliation{INFN, Sezione di Milano Bicocca e Dipartimento di Fisica Università di Milano Bicocca, Italy}

\author{A. Brigatti}
\affiliation{INFN, Sezione di Milano e Università degli Studi di Milano, Dipartimento di Fisica, Italy}

\author{R. Bruno}
\affiliation{INFN, Sezione di Catania e  Università di Catania, Dipartimento di Fisica e Astronomia, Italy}

\author{A. Budano}
\affiliation{INFN, Sezione di Roma Tre e Università degli Studi Roma Tre, Dipartimento di Matematica e Fisica, Italy}

\author{B. Caccianiga}
\affiliation{INFN, Sezione di Milano e Università degli Studi di Milano, Dipartimento di Fisica, Italy}

\author{A. Cammi}
\affiliation{INFN, Sezione di Milano Bicocca e Dipartimento di Energetica, Politecnico di Milano, Italy}

\author{R. Caruso}
\affiliation{INFN, Sezione di Catania e  Università di Catania, Dipartimento di Fisica e Astronomia, Italy}

\author{D. Chiesa}
\affiliation{INFN, Sezione di Milano Bicocca e Dipartimento di Fisica Università di Milano Bicocca, Italy}

\author{C. Clementi}
\affiliation{INFN, Sezione di Perugia e Università degli Studi di Perugia, Dipartimento di Chimica, Biologia e Biotecnologie, Italy}

\author{C. Coletta}
\affiliation{INFN, Sezione di Milano Bicocca e Dipartimento di Fisica Università di Milano Bicocca, Italy}

\author{S. Dusini}
\affiliation{INFN, Sezione di Padova e Università di Padova, Dipartimento di Fisica e Astronomia, Italy}

\author{A. Fabbri}
\affiliation{INFN, Sezione di Roma Tre e Università degli Studi Roma Tre, Dipartimento di Matematica e Fisica, Italy}

\author{G. Felici}
\affiliation{Laboratori Nazionali dell'INFN di Frascati, Italy}

\author{G. Ferrante}
\affiliation{INFN, Sezione di Milano Bicocca e Dipartimento di Fisica Università di Milano Bicocca, Italy}

\author{M.G. Giammarchi}
\affiliation{INFN, Sezione di Milano e Università degli Studi di Milano, Dipartimento di Fisica, Italy}

\author{N. Giudice}
\affiliation{INFN, Sezione di Catania e  Università di Catania, Dipartimento di Fisica e Astronomia, Italy}

\author{N. Guardone}
\affiliation{INFN, Sezione di Catania e  Università di Catania, Dipartimento di Fisica e Astronomia, Italy}

\author{F. Houria}
\affiliation{INFN, Sezione di Milano e Università degli Studi di Milano, Dipartimento di Fisica, Italy}

\author{C. Landini}
\affiliation{INFN, Sezione di Milano e Università degli Studi di Milano, Dipartimento di Fisica, Italy}

\author{I. Lippi}
\affiliation{INFN, Sezione di Padova e Università di Padova, Dipartimento di Fisica e Astronomia, Italy}

\author{L. Loi}
\affiliation{INFN, Sezione di Milano Bicocca e Dipartimento di Energetica, Politecnico di Milano, Italy}

\author{P. Lombardi}
\affiliation{INFN, Sezione di Milano e Università degli Studi di Milano, Dipartimento di Fisica, Italy}

\author{F. Mantovani}
\affiliation{INFN, Sezione di Ferrara, Italy}
\affiliation{Università degli Studi di Ferrara, Dipartimento di Fisica e Scienze della Terra, Italy}

\author{S.M. Mari}
\affiliation{INFN, Sezione di Roma Tre e Università degli Studi Roma Tre, Dipartimento di Matematica e Fisica, Italy}

\author{A. Martini}
\affiliation{Laboratori Nazionali dell'INFN di Frascati, Italy}

\author{L. Miramonti}
\affiliation{INFN, Sezione di Milano e Università degli Studi di Milano, Dipartimento di Fisica, Italy}

\author{M. Montuschi}
\affiliation{INFN, Sezione di Ferrara, Italy}
\affiliation{Università degli Studi di Ferrara, Dipartimento di Fisica e Scienze della Terra, Italy}

\author{M. Nastasi}
\affiliation{INFN, Sezione di Milano Bicocca e Dipartimento di Fisica Università di Milano Bicocca, Italy}

\author{D. Orestano}
\affiliation{INFN, Sezione di Roma Tre e Università degli Studi Roma Tre, Dipartimento di Matematica e Fisica, Italy}

\author{F. Ortica}
\affiliation{INFN, Sezione di Perugia e Università degli Studi di Perugia, Dipartimento di Chimica, Biologia e Biotecnologie, Italy}

\author{A. Paoloni}
\affiliation{Laboratori Nazionali dell'INFN di Frascati, Italy}

\author{L. Pelicci}
\affiliation{INFN, Sezione di Milano e Università degli Studi di Milano, Dipartimento di Fisica, Italy}

\author{E. Percalli}
\affiliation{INFN, Sezione di Milano e Università degli Studi di Milano, Dipartimento di Fisica, Italy}

\author{F. Petrucci}
\affiliation{INFN, Sezione di Roma Tre e Università degli Studi Roma Tre, Dipartimento di Matematica e Fisica, Italy}

\author{E. Previtali}
\affiliation{INFN, Sezione di Milano Bicocca e Dipartimento di Fisica Università di Milano Bicocca, Italy}

\author{G. Ranucci}
\affiliation{INFN, Sezione di Milano e Università degli Studi di Milano, Dipartimento di Fisica, Italy}

\author{A.C. Re}
\affiliation{INFN, Sezione di Milano e Università degli Studi di Milano, Dipartimento di Fisica, Italy}

\author{B. Ricci}
\affiliation{INFN, Sezione di Ferrara, Italy}
\affiliation{Università degli Studi di Ferrara, Dipartimento di Fisica e Scienze della Terra, Italy}

\author{A. Romani}
\affiliation{INFN, Sezione di Perugia e Università degli Studi di Perugia, Dipartimento di Chimica, Biologia e Biotecnologie, Italy}

\author{C. Sirignano}
\affiliation{INFN, Sezione di Padova e Università di Padova, Dipartimento di Fisica e Astronomia, Italy}

\author{M. Sisti}
\affiliation{INFN, Sezione di Milano Bicocca e Dipartimento di Fisica Università di Milano Bicocca, Italy}

\author{L. Stanco}
\affiliation{INFN, Sezione di Padova e Università di Padova, Dipartimento di Fisica e Astronomia, Italy}

\author{E. Stanescu Farilla}
\affiliation{INFN, Sezione di Roma Tre e Università degli Studi Roma Tre, Dipartimento di Matematica e Fisica, Italy}

\author{V. Strati}
\affiliation{INFN, Sezione di Ferrara, Italy}
\affiliation{Università degli Studi di Ferrara, Dipartimento di Fisica e Scienze della Terra, Italy}

\author{M.D.C Torri}
\affiliation{INFN, Sezione di Milano e Università degli Studi di Milano, Dipartimento di Fisica, Italy}

\author{C. Tuvè}
\affiliation{INFN, Sezione di Catania e  Università di Catania, Dipartimento di Fisica e Astronomia, Italy}

\author{C. Venettacci}
\affiliation{INFN, Sezione di Roma Tre e Università degli Studi Roma Tre, Dipartimento di Matematica e Fisica, Italy}

\author{G. Verde}
\affiliation{INFN, Sezione di Catania e  Università di Catania, Dipartimento di Fisica e Astronomia, Italy}

\author{L. Votano}
\affiliation{Laboratori Nazionali dell'INFN di Frascati, Italy}

\begin{abstract}
{Precise modeling of detector energy response is crucial for next-generation neutrino experiments which present computational challenges due to lack of analytical likelihoods. We propose a solution using neural likelihood estimation within the simulation-based inference framework. We develop two complementary neural density estimators that model likelihoods of calibration data: conditional normalizing flows and a transformer-based regressor. We adopt JUNO --- a large neutrino experiment --- as a case study. The energy response of JUNO depends on several parameters, all of which should be tuned, given their non-linear behavior and strong correlations in the calibration data. To this end, we integrate the modeled likelihoods with Bayesian nested sampling for parameter inference, achieving uncertainties limited only by statistics with near-zero systematic biases. The normalizing flows model enables unbinned likelihood analysis, while the transformer provides an efficient binned alternative. By providing both options, our framework offers flexibility to choose the most appropriate method for specific needs. Finally, our approach establishes a template for similar applications across experimental neutrino and broader particle physics.}
\end{abstract}


\maketitle

\section{Introduction}
\label{sec:intro}

The new precision era in neutrino physics~\cite{Capozzi:2025wyn} demands accurate detector energy response characterization. The relationship between the energy released by a neutrino interaction in the detector (deposited energy) and the measured experimental quantities rarely follows a simple analytical form in high-precision detectors, instead representing a complex function, dependent on detector geometry, material properties, and interacting particles.

The absence of analytical descriptions necessitates sophisticated Monte Carlo (MC) simulations that account for a wide range of physics effects. However, these simulations may depend on multiple parameters that summarize complex behaviors. Despite implementing detector properties based on the best available knowledge, inherent uncertainties and necessary simplifications remain. These limitations mean that simulations typically require further refinement to accurately reproduce experimental observations through a process known as MC tuning, which involves systematic adjustment of simulation parameters to achieve an agreement between simulated outputs and experimental data.

For MC tuning, parameter inference requires a likelihood function $p(x | \bm{\phi})$, where $x$ represents either simulated or observed data and $\bm{\phi}$ are the parameters to be tuned. Once constructed, this likelihood enables recovery of the parameter values $\bm{\phi}$ that best match the observed data $x$. However, this likelihood function is computationally intractable directly from simulations. Approaches employed in neutrino experiments of the previous generation, like Borexino~\cite{Agostini2018} and Daya Bay~\cite{DayaBay:2016} involve iterative MC tuning where parameters are adjusted manually to match calibration data. Yet this becomes computationally prohibitive for modern experiments.

We employ the Neural Likelihood Estimation (NLE) method that constructs approximations to the $p(x| \bm{\phi})$ function to address the challenge of the MC tuning. NLE belongs to the broader family of Simulation-Based Inference (SBI, or likelihood-free inference)~\cite{wood2010statistical, Papamakarios:2016ctj, lueckmann2019likelihood, papamakarios2019sequential, Cranmer:2019eaq, Lueckmann:2021}, now widely used in multiple scientific domains, including cosmology~\cite{Alsing:2018, Alsing:2019, Kosiba:2024qku}, astrophysics~\cite{Brehmer:2019jyt, Dax:2021tsq, DES:2024xij}, accelerator~\cite{Brehmer:2019xox, Lazzarin:2020uvv, Heller:2024onk, ATLAS:2025clx} and neutrino physics~\cite{Pina-Otey:2020tdz, ElBaz:2023ijr}, and beyond~\cite{Brehmer:2018eca, Brehmer:2018kdj}. 

Within SBI, there also exists a second strategy called Neural Posterior Estimation (NPE), that learns the posterior $q(\bm{\phi}|x)$ directly and is attractive for ultra-fast inference. NLE, by contrast, separates the likelihood $p(x|\bm{\phi})$ estimation from statistical inference. In our implementation, we first train a model for the likelihood estimation using simulation data and then combine that model with conventional statistical tools to obtain the posterior. This division retains full control over statistical interpretation and uncertainty quantification. The choice between NPE and NLE is guided by data dimensionality, simulation cost, and the need for transparent statistical treatment.

As a case study for demonstrating our methodology, we use the Jiangmen Underground Neutrino Observatory (JUNO)~\cite{JUNO:2015zny, JUNO:2021vlw}. The experiment consists of a central detector containing a $2 \cdot 10^{7}$ kg liquid scintillator-based (LS) target within a 35.4~m diameter acrylic sphere, monitored by 17,612 20'' and 25,600 3'' photomultiplier tubes (PMTs), converting light produced in the detector medium to electric charge. This design yields exceptional light collection, quantified by the total number of photons converted to electrons by the PMTs, i.e., number of photo-electrons $N_{p.e.}$. In our analysis, we exploit only the 20'' PMTs, which provide the most precise energy measurement by detecting on average 1,600 photo-electrons when a 1~MeV gamma loses all its energy at the center of the detector. However, the same approach could be applied also incorporating the 3'' PMTs if needed. JUNO's primary goals include determining the neutrino mass ordering~\cite{JUNO:2024jaw} and measuring three oscillation parameters ($\Delta m^2_{21}$, $\Delta m^2_{31}$, $\sin^2\theta_{12}$) with sub-percent precision~\cite{JUNO:2022mxj}, alongside a broad program exploring phenomena from geoneutrinos to supernovae~\cite{JUNO:2023zty, JUNO:2022jkf, JUNO:2023dnp, JUNO:2022lpc, JUNO:2021tll, JUNO:2022qgr, JUNO:2024pur, JUNO:2023vyz}. As a large-scale, next-generation LS detector, it requires unprecedented energy resolution and control over energy-related systematic uncertainties~\cite{JUNO:2021vlw} to achieve its goals. Thus, JUNO's complex, non-uniform, and non-linear response~\cite{JUNO:2024fdc}, coupled with its strong reliance on highly precise MC simulations, makes it a suitable platform for developing and validating our approaches.

The energy response in JUNO is characterized by three key parameters with different physical origins and effects: (1) the Birks' coefficient ($k_B$), which models the non-linear energy quenching at high ionization densities, where part of the energy deposited by the particle is dissipated as heat in the LS rather than being converted into scintillation light; (2) the LS absolute light yield ($Y$), which defines the number of scintillation photons emitted per unit energy after quenching; (3) the Cherenkov light yield factor ($f_C$)  which scales the energy-dependent yield of photons originating from Cherenkov radiation. Both $k_B$ and $f_C$ model the non-linear relation between the deposited energy and the collected charge, while $Y$ scales energy response linearly~\cite{JUNO:2024fdc}.

Precisely calibrating these parameters within JUNO's MC simulation framework~\cite{JUNO:juno_soft} is pivotal for mitigating systematic uncertainties. Furthermore, accurate simulations are important for any ML applications used in JUNO~\cite{Qian:2021vnh, Gavrikov:2022, Yang:2023rbg, Gavrikov:2024rso, Jiang:2024wph, Malyshkin:chep2024, liu2025neutrino}, preventing performance degradation due to discrepancies between simulated training data and real experimental data.

To constrain the energy response model and determine the optimal ($k_B, f_C, Y$) parameters, JUNO employs a comprehensive calibration program using radioactive sources deployed at various positions within the detector~\cite{JUNO:2020xtj}. The strong correlations between $k_B$, $f_C$, and $Y$ require a combined analysis of multiple sources with different energy spectra to break parameter degeneracies. In this work, we use simulated data corresponding to five different calibration sources placed at the detector center: three gamma sources (${}^{137}$Cs, ${}^{40}$K, ${}^{60}$Co) and two neutron sources (${}^{241}$Am-Be, ${}^{241}$Am-${}^{13}$C). For neutron sources, the final state usually consists of an excited state (${}^{12}$C$^*$, ${}^{16}$O$^*$, respectively) emitting a gamma, and a neutron, which is captured by hydrogen (n-H) or carbon (n-$^{12}$C) atoms in the LS, leading to the emission of an additional gamma. Further details on these sources are available in Ref.~\cite{JUNO:2020xtj}. 

For each event of the calibration data, the total signal collected by the PMT array is summarized by the total number of photo-electrons $N_{p.e.}$ detected by all channels. Since the energy response is modeled at the event level, this compression of the full PMT-wise information into one-dimensional $N_{p.e.}$ distributions enables efficient likelihood modeling while preserving the information relevant for parameter estimation. $N_{p.e.}$ serves as a proxy for the visible energy of an event and translates to the $x$ in the notation used above, while $\bm{\phi}$ is the vector of the energy response parameters ($k_B, f_C, Y$).

We develop two complementary neural density estimators that learn to approximate the likelihood function from calibration data: a Transformer Encoder Density Estimator (TEDE)~\cite{transformer} that maps energy response parameters to histogram-based density approximations, and a Normalizing Flows Density Estimator (NFDE)~\cite{tabak_nf, rezende2015variational, trippe2018conditional} that models continuous probability densities for unbinned analysis. We integrate these learned likelihood approximations with Bayesian nested sampling~\cite{ultranest, buchner2023nested} for inference of the parameters $\bm{\phi}$. This paper also represents a substantial extension of our previous study available in Ref.~\cite{neuro_mct_chep}, where we studied TEDE and a Wasserstein generative adversarial network  with gradient penalty~\cite{gan, wgan-gp}.

Additionally, we provide a comprehensive systematic uncertainty estimation analysis, demonstrating that both methods achieve unbiased parameter estimation with uncertainties limited only by statistics of the calibration data.

In summary, both developed NLE methods, TEDE and NFDE, integrated with the Bayesian nested sampling technique successfully recover JUNO's energy response parameters ($k_B$, $f_C$, $Y$). Through our systematic uncertainty estimation analysis across 1,000 parameter combinations and varying statistical exposures, we demonstrate robust parameter estimation with proper uncertainty quantification in a highly correlated parameter space. We show that both methods provides the parameter estimates with near-zero systematic bias and uncertainties limited only by statistics. Beyond the application to JUNO, this work establishes a methodological template for applying NLE across experimental neutrino physics for various problems where analytical likelihoods are intractable. As experiments like DUNE~\cite{DUNE:2020} and Hyper-Kamiokande~\cite{HyperK:2018} come online with increasingly stringent precision requirements, NLE (and SBI methods, in general) will become essential tools for extracting maximum physics information from data.

\section{Results}
\label{sec:results}

As introduced in~\autoref{sec:intro}, the NLE approach requires models that can approximate the conditional probability density function $p(x | \bm{\phi})$, where $x$ represents the observations (the total number of photo-electrons $N_{p.e.}$ as a proxy to the visible energy) and $\bm{\phi}$ denotes the parameters to be inferred. We present two complementary NLE implementations: TEDE and NFDE, both targeting the conditional density $p(x | \bm{\psi})$, where $\bm{\psi} = \bm{\phi} \cup S = \{k_B, f_C, Y, S\}$ combines the energy response parameters ($\bm{\phi}$) with the calibration source type ($S$). Incorporating $S$ as an additional conditioning variable alongside $\bm{\phi}$ enables us to utilize models in a unified way to handle different calibration sources during the parameter inference process. Thus, the TEDE and NFDE models serve as fast likelihood estimators for the computationally expensive MC simulations.

While both approaches target the same conditional density estimation task, they differ in how they handle the problem. In the settings of the TEDE approach, we first approximate the conditional density using a histogram representation, then the TEDE model is trained to learn a direct mapping from $\bm{\psi}$ to the bin-to-bin approximations of the corresponding density. This approach naturally integrates with binned likelihood analyses. In contrast, NFDE directly models the continuous probability density through normalizing flows, enabling unbinned likelihood evaluation that preserves the full information content of the data. Below, we briefly outline the technical details of both models, while their comprehensive description can be found in~\autoref{subsec:ml_models}. Additionally, model training and hyperparameter optimization are described in~\autoref{subsec:hopt_and_training}.

Specifically, the TEDE model uses a transformer encoder architecture to learn the mapping from input parameters $\bm{\psi}$ to histogram-based density approximations. The energy range, expressed in $N_{p.e.}$, from 400 to 16,400, is divided into $N_b=800$ bins with a width of $\Delta x = 20 \, N_{p.e.}$. The model produces probability density values for each bin through a temperature-scaled softmax layer, which is given by:
\begin{equation}
\left\{\hat{p}^{\rm TEDE}_i \right\}_{i=1 \dots N_b} = \frac{1}{\Delta x} \left\{ \frac{e^{z_i/T}}{\sum_{j=1}^{N_b} e^{z_j/T}} \right\}_{i=1 \dots N_b},
\end{equation}
where $z_i$ are the raw values from the proceeding layer and $T$ is the softmax temperature hyperparameter, which controls the sharpness of the output distribution. By construction, this ensures proper normalization of the probability distribution function (PDF): $\sum_{i=1}^{N_b}\hat{p}_i^{\rm TEDE} \Delta x = 1$.

The NFDE model employs an exact approach to conditional density estimation based on normalizing flows. In the density estimation setting, the normalizing flows model is trained to transform a complex distribution into a well-known simple \textit{base} distribution through a sequence of invertible transformations, \textit{flows}. Usually, and in our specific case, the normal distribution is used as the base distribution. Invertibility of the flows allows transformation in both directions. The direction from the complex distribution to the base distribution is called \textit{normalizing direction} (following the fact of normal distribution to be a common choice), while the opposite direction is called \textit{generative direction}, since it allows to produce new samples of the complex distribution by sampling from the base and passing the base samples through the model. The density estimation is based on the normalizing direction of the flow because it allows the exact likelihood computation for each value of the complex distribution. 
We train the NFDE model to transform the complex energy distributions of the calibration sources to the standard normal distribution through a sequence of invertible transformations $\bm{f}_{\bm{\theta}}$, conditioned on $\bm{\psi}$ via dedicated neural network-based parameterizations. The exact probability density at any energy value $x$ is then computed using the change of variables formula:
\begin{equation}
\hat{p}(x|\bm{\psi}) = p_Z(\bm{f}_{\bm{\theta}}(x|\bm{\psi})) \left| \frac{\partial \bm{f}_{\bm{\theta}}(x|\bm{\psi})}{\partial x} \right|,
\end{equation}
where $p_Z$ is the density of the standard normal distribution and the Jacobian determinant accounts for the volume change induced by the transformation.

\begin{figure}[!htb]
    \centering
    \includegraphics[width=0.49\textwidth]{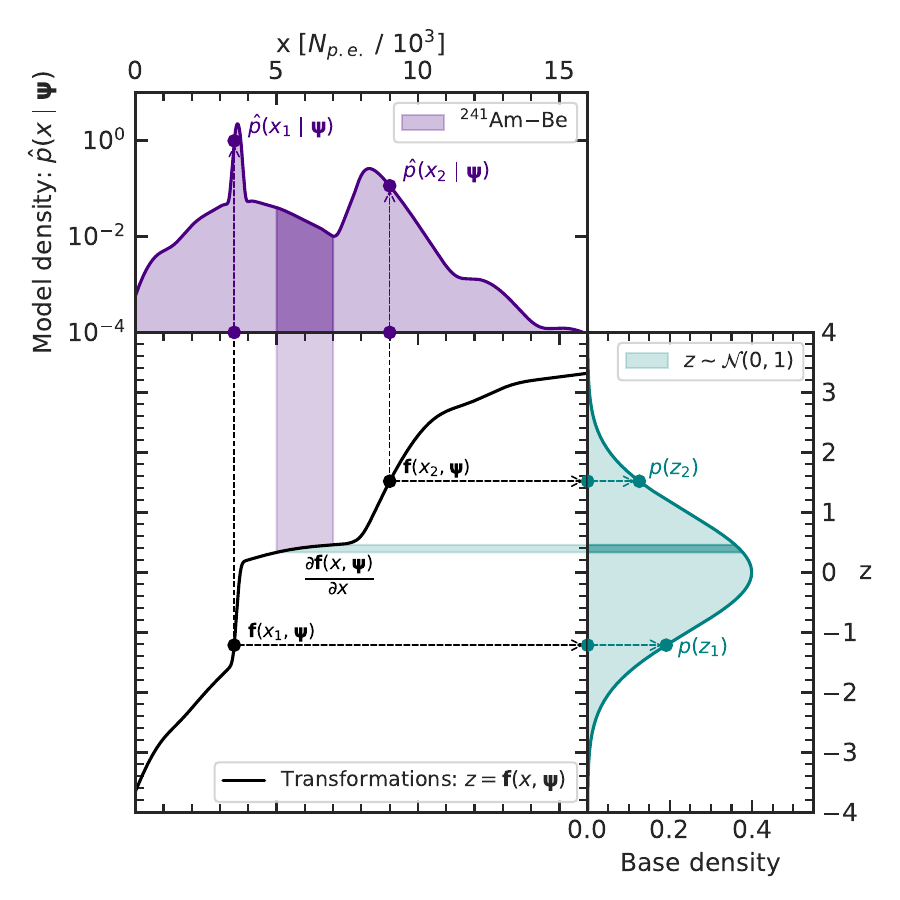}
    \caption{\textbf{Illustration of the normalizing direction of the NFDE model.} The plot presents how complex energy distribution of ${}^{241}$Am-Be is transformed into a simple Gaussian distribution through a series of learned invertible mappings $\bm{f}_{\bm{\theta}} \colon z = \bm{f}_{\bm{\theta}}(x, \bm{\psi}) = f_K \left(f_{K-1} \left(\dots \ f_1 \left(x, \bm{\psi}\right) \right) \right)$, where $\bm{\psi} = \{k_B, f_C, Y, S\}$ is the vector of conditioning variables and $K$ is the number of flows. More specifically, the NFDE model transforms the values $x$ (the visible energy expressed in $N_{p.e.}$, top left) to latent space points $z$ (bottom right) through sequential flow transformations (center). The conditional probability density $\hat{p}(x | \bm{\psi})$ for any value of $x$ given parameters $\bm{\psi}$ is calculated by applying the change-of-variables formula: $\hat{p}(x | \bm{\psi}) = p_Z(z) \cdot \left|{\partial \bm{f}_{\bm{\theta}}(x,\bm{\psi})}/{\partial x}\right|$, where $p_Z(z)$ is the standard normal density and ${\partial \bm{f}_{\bm{\theta}}(x,\bm{\psi})}/{\partial x}$ is the Jacobian of the composite transformation $\bm{f}_{\bm{\theta}}$. This approach enables exact likelihood evaluation needed for the unbinned parameter estimation procedure.}
    \label{fig:nfde_inference_vis}
\end{figure}

The principle of this normalizing flow approach is illustrated in \autoref{fig:nfde_inference_vis}, which shows how the complex, multimodal energy distribution of ${}^{241}$Am-Be is transformed through a sequence of conditional flows into a simple Gaussian distribution. This transformation enables exact likelihood computation for unbinned data points --- a key advantage for parameter estimation.

We evaluate the performance of both approaches in two stages. First, we demonstrate the ability of the models to accurately reproduce complex energy spectra across the parameter space. Then, we assess their effectiveness when integrated into the parameter estimation framework outlined in~\autoref{subsec:parameter_estimation}, evaluating the accuracy, precision, and robustness of the resulting energy response parameter estimations.

\begin{figure*}[!htb]
	\centering
	\includegraphics[width=1\textwidth]{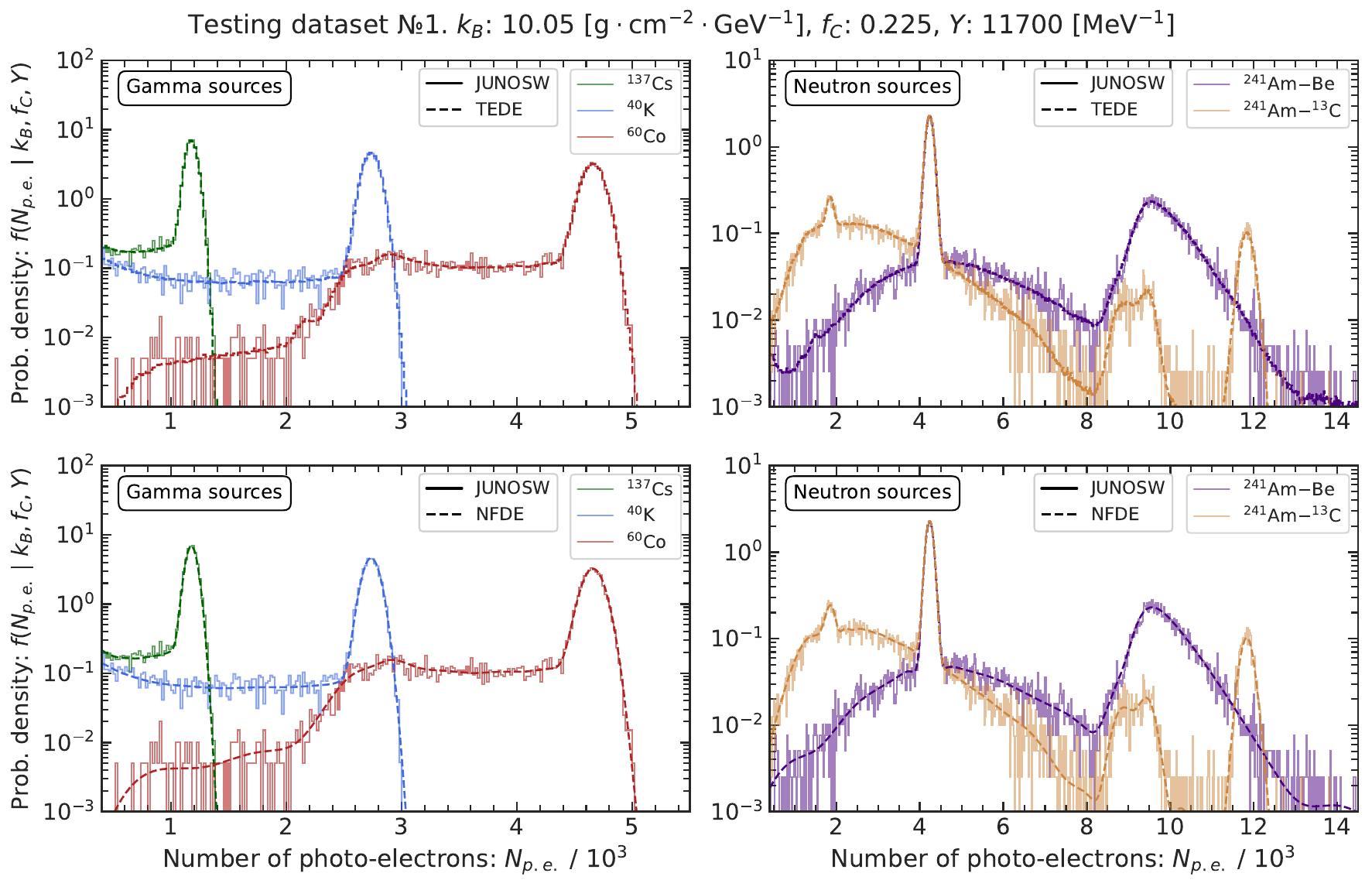}
	\caption{\textbf{Comparison of the modeled PDFs with the corresponding energy spectra.} All five calibration sources are shown with the following values of the parameters: $k_B = 10.05 \text{ g}\cdot \text{cm}^{-2} \cdot \text{GeV}^{-1}$, $f_C = 0.225$, $Y = 11700 \text{ MeV}^{-1}$. Left panel shows the gamma sources and the right panel shows the neutron sources. The separation between the types of the sources is for visualization purposes only. The top and the bottom row present the modeled energy spectrum of each source built using TEDE and NFDE, respectively. Here, JUNOSW stands for ``JUNO software''. Both models show high ability to accurately reproduce the expected energy distributions of all sources, while NFDE presents smooth continuous curves.}
\label{fig:models_comparison_spectra}
\end{figure*}

Our study is based on extensive Monte Carlo data produced using the official JUNO software~\cite{JUNO:juno_soft}, containing approximately 1 billion simulated events across five calibration sources ($^{137}$Cs, $^{40}$K, $^{60}$Co, $^{241}$Am-Be, and $^{241}$Am-$^{13}$C) for different combinations of the parameters. Here, each event is ultimately represented by a single value of $N_{p.e.}$, which is conditioned on the parameters $\bm{\psi}$. Following standard machine learning practices, we organize data into separate training ($\sim$600 million events), validation ($\sim$270 million events), and testing ($\sim$370 million events) datasets, with specific parameter grids designed to evaluate model performance in an unbiased manner. The training dataset covers a comprehensive grid of 9,261 parameter combinations (21 grid points per each parameter), while validation and testing datasets are positioned exactly in the middle between the training points to obtain models' results in the worst case scenario. We produce two types of validation datasets: validation dataset 1 covering a grid of 1,000 parameter points and validation dataset 2 with high statistics at three specific points. Similarly, we construct two testing datasets: a testing dataset 1 with 1,000 different parameter combinations (shifted with respect to both the validation grid and the training grid) and a testing dataset 2 with a focus on a single parameter point (close to the center of the grid) with varying statistical exposures to perform a systematic uncertainty analysis. A detailed description of these datasets is provided in~\autoref{subsec:data_description}.

\subsection{Models performance}
\label{sec:results:models_performance}

\begin{figure*}[!htb]
	\centering
	\includegraphics[width=1\textwidth]{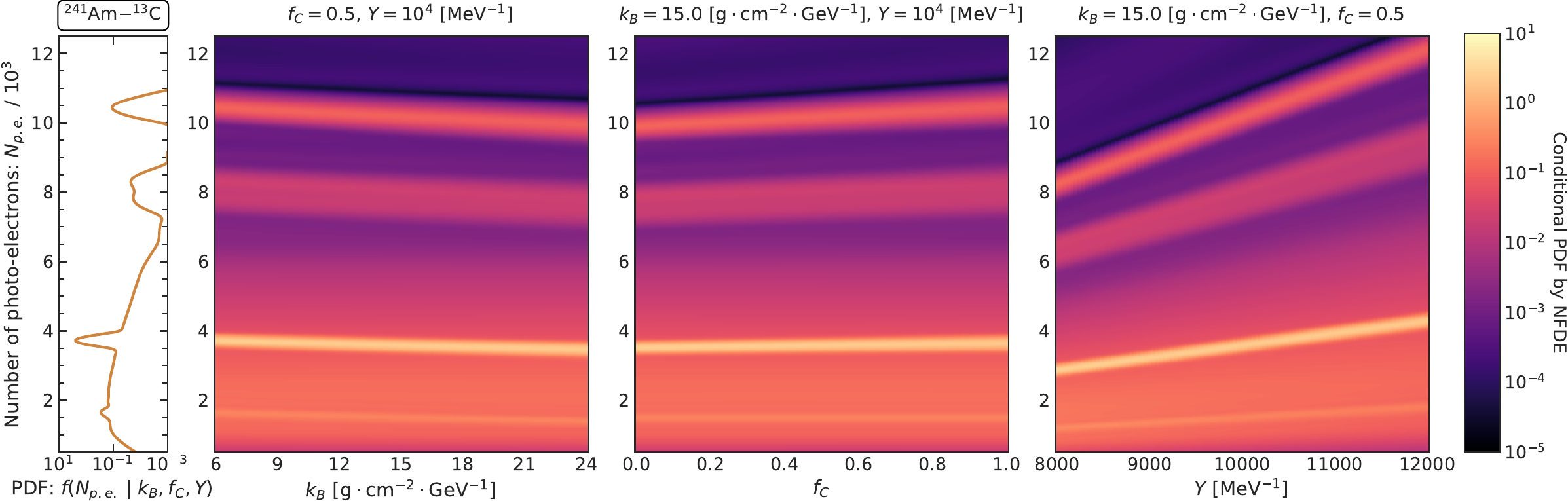}
    \caption{\textbf{Visualization of how the energy response parameters impact the calibration source energy spectra, shown through conditional PDF modeled by NFDE.} Heatmaps present the probability density given by the NFDE model for the ${}^{241}$Am-${}^{13}$C neutron source as each parameter is varied independently: $k_B$ (left), $f_C$ (middle), and $Y$ (right). The rotated spectrum on the left shows the modeled by NFDE spectrum of ${}^{241}$Am-${}^{13}$C for illustration; the following parameter values were used: $k_B = 6.0 \text{ g}\cdot \text{cm}^{-2} \cdot \text{GeV}^{-1}$, $f_C = 0.5$, $Y = 10000 \text{ MeV}^{-1}$. For each panel, the two parameters that are not varied are fixed at their median grid values. Color intensity represents conditional PDF values on a logarithmic scale, showing the expected parameter-specific effects. Note, that the y-axis upper limit of 12.5k $N_{p.e.}$ is chosen for visualization purposes only, as it encompasses the main parts of the spectra. Similar plot for the TEDE model can be found in~\autoref{fig:conditional_pdf_heatmap_tede_AmC}.}
\label{fig:conditional_pdf_heatmap_nfde_AmC}
\end{figure*}

To evaluate and compare the ability of both models to properly describe the data we use a family of statistical distances based on the $L_p$-norm framework, as detailed in~\autoref{subsec:metrics}. These metrics --- Wasserstein ($p=1$), Cramér-von Mises ($p=2$), and Kolmogorov-Smirnov ($p=\infty$) distances --- quantify different aspects of the discrepancy between predicted and true probability distributions. Among these, we selected the Cramér-von Mises distance ($d_2$) as our primary optimization criterion due to its balanced sensitivity to both global distribution shifts and local deformations, making it particularly suitable for evaluating complex energy spectra with diverse features.

\autoref{fig:models_comparison_spectra} provides a visual comparison between the PDFs predicted by both models and the corresponding ground truth empirical PDF (limited in statistics) for all five calibration sources at a representative point in the parameter space ($k_B = 10.05 \text{ g}\cdot \text{cm}^{-2} \cdot \text{GeV}^{-1}$, $f_C = 0.225$, $Y = 11{,}700 \text{ MeV}^{-1}$). Both models demonstrate high fidelity in capturing diverse spectral features, including sharp peaks, Compton edges, and the continuous components characteristic of different source types. The TEDE model produces a binned approximation that closely follows the target distribution while showing some insignificant residual fluctuations in the low statistics regions, while the NFDE model generates smooth continuous curves that accurately represent the spectra PDFs, reproducing the very tiny spectrum features (such as the double gamma peak of the $^{241}$Am-$^{13}$C at around 9,000 $N_{p.e.}$).

To quantitatively assess model performance,~\autoref{tab:model_performance_metrics} presents a comparison of the statistical distance metrics for both models across validation and testing datasets. Both models achieve similar near-zero discrepancies with the data PDFs, validating both approaches as effective one-dimensional conditional density estimators. However, TEDE slightly outperforms NFDE in all the metrics with differences within the same order of magnitude. Interestingly, while NFDE shows almost the same performance on the primary Cramér-von Mises ($d_2$) metric, it demonstrates noticeably worse results on Kolmogorov-Smirnov metric. This likely reflects the KS distance's higher sensitivity to localized distribution differences and outliers, as it measures maximum deviation between cumulative distribution functions (CDFs) rather than integrated differences across the entire distribution. It is worth noting that when comparing these metrics, there exists an additional source of uncertainty related to the methodological differences in CDF estimation between models: TEDE relies on binned histogram approximations, while NFDE employs unbinned analytical CDFs. This methodological difference may introduce small systematic variations in the distance measurements that are independent of the actual model performance.

\begin{table}[!htb]
\centering
\setlength{\tabcolsep}{6pt}
\begin{tabular}{|l|cc|cc|}
\hline
\multirow{2}{*}{Statistical distance} & \multicolumn{2}{c|}{Validation dataset} & \multicolumn{2}{c|}{Testing dataset} \\
\cline{2-5}
& TEDE & NFDE & TEDE & NFDE \\
\hline
$d_1 \times 10^{-3}$ & 5.0 & 5.7 & 4.5 & 5.4 \\
$d_2 \times 10^{-3}$ & 2.6 & 2.6 & 2.4 & 2.6 \\
$d_\infty \times 10^{-3}$ & 3.8 & 4.4 & 3.5 & 5.6 \\
\hline
\end{tabular}
\caption{\textbf{Performance comparison of TEDE and NFDE models.} The table presents three statistical distance for both validation and testing datasets used as metrics in this study. Here, $d_1$, $d_2$, $d_\infty$ are the Wasserstein, Cramér-von Mises and Kolmogorov-Smirnov distances, respectively. The metrics values represent averages between the broad grid but sparse-statistics dataset (validation dataset 1) and three points high-statistics dataset (validation dataset 2), with similar averaging for the corresponding testing datasets. All values are scaled by $10^{-3}$ for readability. Lower values indicate better model performance in approximating the true probability distributions.}
\label{tab:model_performance_metrics}
\end{table}

Beyond accurately modeling individual spectra, the models must correctly capture the functional dependence of the spectrum shape on the energy response parameters. \autoref{fig:conditional_pdf_heatmap_nfde_AmC} visualizes this capability for the NFDE model using the ${}^{241}$Am-${}^{13}$C source as an example. The heatmaps illustrate how the predicted conditional PDF evolves as each parameter is varied independently while holding the others fixed. 
Notably, the behavior of the modeled PDF varies smoothly and continuously across the parameter space, with no artificial discontinuities, demonstrating that the NFDE has learned physically consistent parameter dependencies essential for reliable interpolation during parameter estimation.

\begin{figure*}[!htb]
    \centering
    \includegraphics[width=1\textwidth]{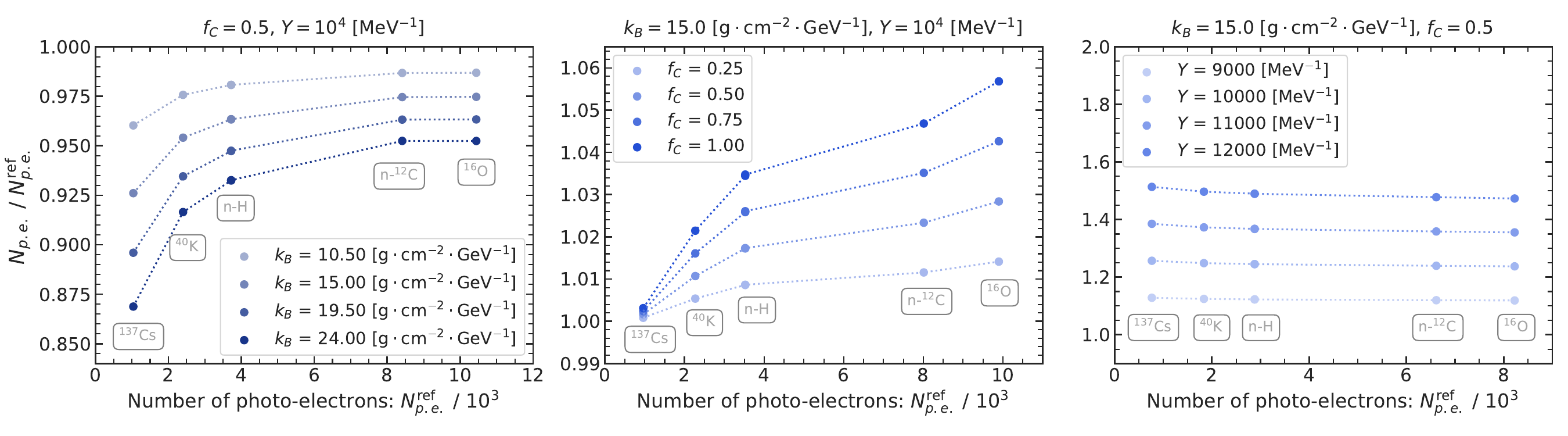}
    \caption{\textbf{The effect of the parameters as a function of energy.} This plot demonstrates how each parameter affects the relative positions of the main peaks across the energy scale of the calibration sources. The main peaks of the spectra from multiple calibration sources serve as standard candles: $^{137}$Cs, $^{40}$K, and neutron capture peaks (n-H, n-$^{12}$C) as well as the $^{16}$O excitation. Moving along the y-axis shows the same gamma peak under different energy response scenarios (different parameter values), while moving along the x-axis shows gamma peaks (i.e., different energies) under the same energy response scenario. For each parameter ($k_B$, $f_C$, and $Y$), five values spanning the full parameter range were selected while keeping the other two parameters fixed at the mean values of the grid. The y-axis shows the relative shift in peak positions compared to references. The plot also highlights the different influence of each parameter: $k_B$ shows increasing non-linearity at lower energies, $f_C$ exhibits greater effect at higher energies, while $Y$ demonstrates a nearly constant scaling with slight deviations at low energies due to PMT dark noise contributions. Similar plot for the TEDE model can be found in~\autoref{fig:nl_tede}.}
        
    \label{fig:nl_nfde}
\end{figure*}

While the $k_B$ and $f_C$ parameters introduce a non-linear relationship between deposited energy of an event and the collected charge (in $N_{p.e.}$), the dependence of the model's output PDF values on the parameters for a fixed deposited energy may or may not be linear. For example, in \autoref{fig:conditional_pdf_heatmap_nfde_AmC}, the main gamma peaks shift as each parameter varies. For the $k_B$ parameter, this shift is roughly linear, which is consistent with the first-order Taylor expansion of Birks' law being linear in $k_B$. For the $f_C$ parameter, the behavior is more complex: there is almost no effect on the lower energy peaks, while the higher energy peaks show a more pronounced, approximately linear dependence on the parameter value. Additionally, the $f_C$ parameter causes the peaks to become wider as the parameter increases, indicating a degradation in energy resolution. Finally, the dependence learned for the $Y$ parameter is truly linear by the nature of the parameter. This dependence of the PDF on the parameters should not be confused with the role that they have in creating a non-linear energy response across the energy scale for a given set of parameter values.

To analyze the non-linear relationship of energy response across different parameter values shown in~\autoref{fig:nl_nfde}, we use the following approach. First, we select the main peaks of the spectra as standard candles: $^{137}$Cs, $^{40}$K, n-H, n-$^{12}$C peaks from neutron sources as well as the $^{16}$O of the $^{241}$Am-$^{13}$C source. Note that we exclude the $^{60}$Co photo-peak and peaks from positron annihilation as they originate from the combined energy deposition of two simultaneous gamma emissions, thus effectively appearing more quenched.
For each parameter, we take five values spanning the full range of the grid while keeping the other two parameters fixed at the mean values of the grid. We use spectra with the lowest parameter values as references. Using the NFDE model, we produce the PDFs for each parameter combination and measure the relative positions (in the number of photo-electrons) of each peak compared to the reference spectra.
This analysis reveals different magnitudes of the parameter effects as shown in~\autoref{fig:nl_nfde}. Light yield ($Y$) demonstrates the strongest overall influence, followed by Birks' constant ($k_B$) and then by the Cherenkov light yield factor ($f_C$). The effect of $k_B$ is distinctly non-linear, becoming more pronounced at lower energies. Conversely, $f_C$ shows a non-linear behavior with its effect magnitude increasing at higher energies while having minimal impact in the low-energy region and almost zero impact for the $^{137}$Cs source. This makes the $^{137}$Cs source particularly valuable for constraining the other parameters. The light yield response appears nearly constant, with slight deviations at low energies attributable to PMT dark noise contributions, whose relative impact decreases as energy increases.

Though relatively subtle, these non-linear effects are critical for the high precision determination of the detector energy scale (as in the case of JUNO~\cite{JUNO:2024fdc}). Using the TEDE model, we obtained similar results to the ones presented for NFDE on ~\autoref{fig:conditional_pdf_heatmap_nfde_AmC} and~\autoref{fig:nl_nfde}, as can be found in~\autoref{ax:tede_param_dependencies}.

The models' ability to learn these complex dependencies (including their non-linear nature) directly from the data is essential for their effectiveness in the subsequent parameter estimation task, particularly for breaking degeneracies between highly correlated parameters. As evidenced by the parameter recovery accuracy shown in subsequent sections, both models successfully captures these small but systematic non-linear effects.

\begin{figure*}[!htb]
	\centering
	\includegraphics[width=1\textwidth]{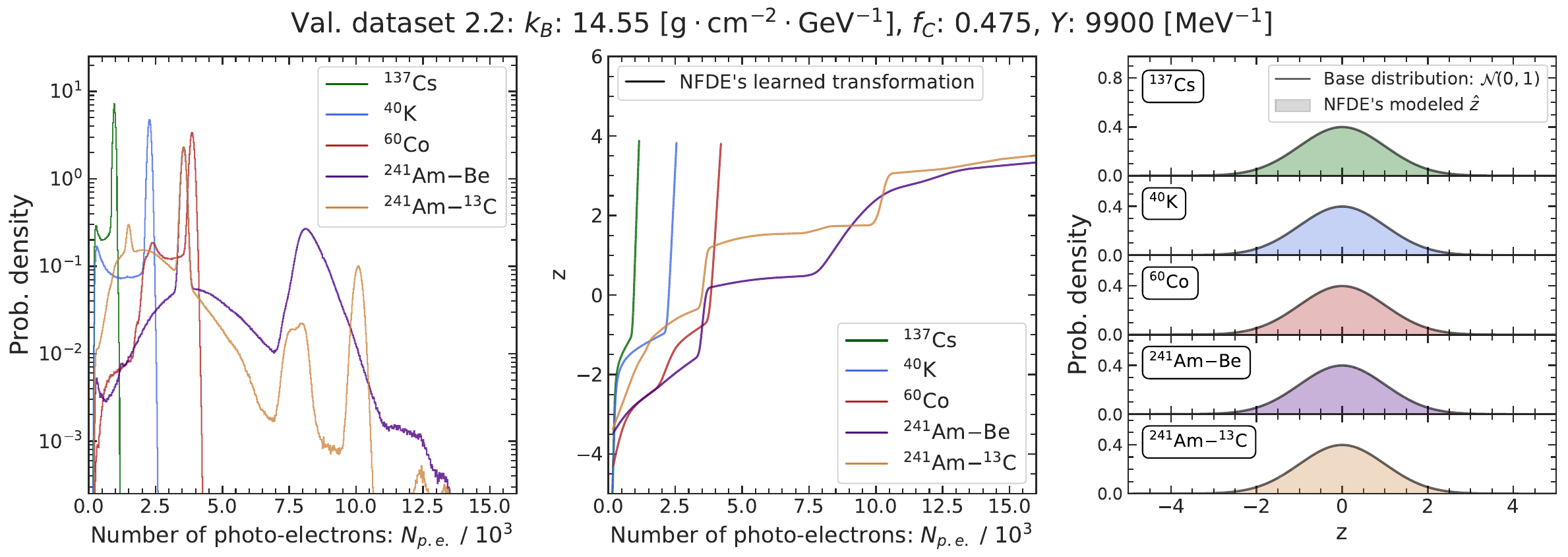}
	\caption{\textbf{Visualization of the normalizing transformations of the NFDE model applied to validation dataset 2.} We present the second point of validation dataset 2 with the following energy response parameters values: $k_B = 14.55 \text{ g}\cdot \text{cm}^{-2} \cdot \text{GeV}^{-1}$, $f_C = 0.475$, $Y = 9900 \text{ MeV}^{-1}$. The plot presents how the NFDE model maps complex energy spectra of the five calibration sources (left panel) through learned invertible transformations (middle panel) to standard normal distributions $\mathcal{N}(0,1)$ (right panel). Different calibration sources are shown in different colors. The results confirm the model's ability to transform diverse multimodal energy distributions into simple unimodal latent representations, while maintaining the conditional information. The successful normalization across all calibration sources demonstrates the NFDE model as a unified approach to handle the complex non-linear detector response across different energy ranges and interaction types.}
    \label{fig:nfde_transformations_val2}
\end{figure*}

The effectiveness of the NFDE approach can be further assessed by examining how well it performs its core function: transforming complex data distributions into the standard normal distribution.~\autoref{fig:nfde_transformations_val2} shows an example of the model's success in this task for one of the points of validation dataset 2. The visualization shows how the complex, multimodal energy distributions of the five calibration sources (left panel) are progressively transformed through the learned flows (middle panel) into well-formed standard normal distributions (right panel). This successful normalization across different source types is persistent across all the parameter space of the validation datasets and thus confirms that the model has learned appropriate conditional transformations for all calibration sources within the parameter space.

\subsection{Parameter estimation}
\subsubsection{Single representative point in the parameter space}

\begin{figure*}[!htb]
  \centering
  \includegraphics[width=0.49\textwidth]{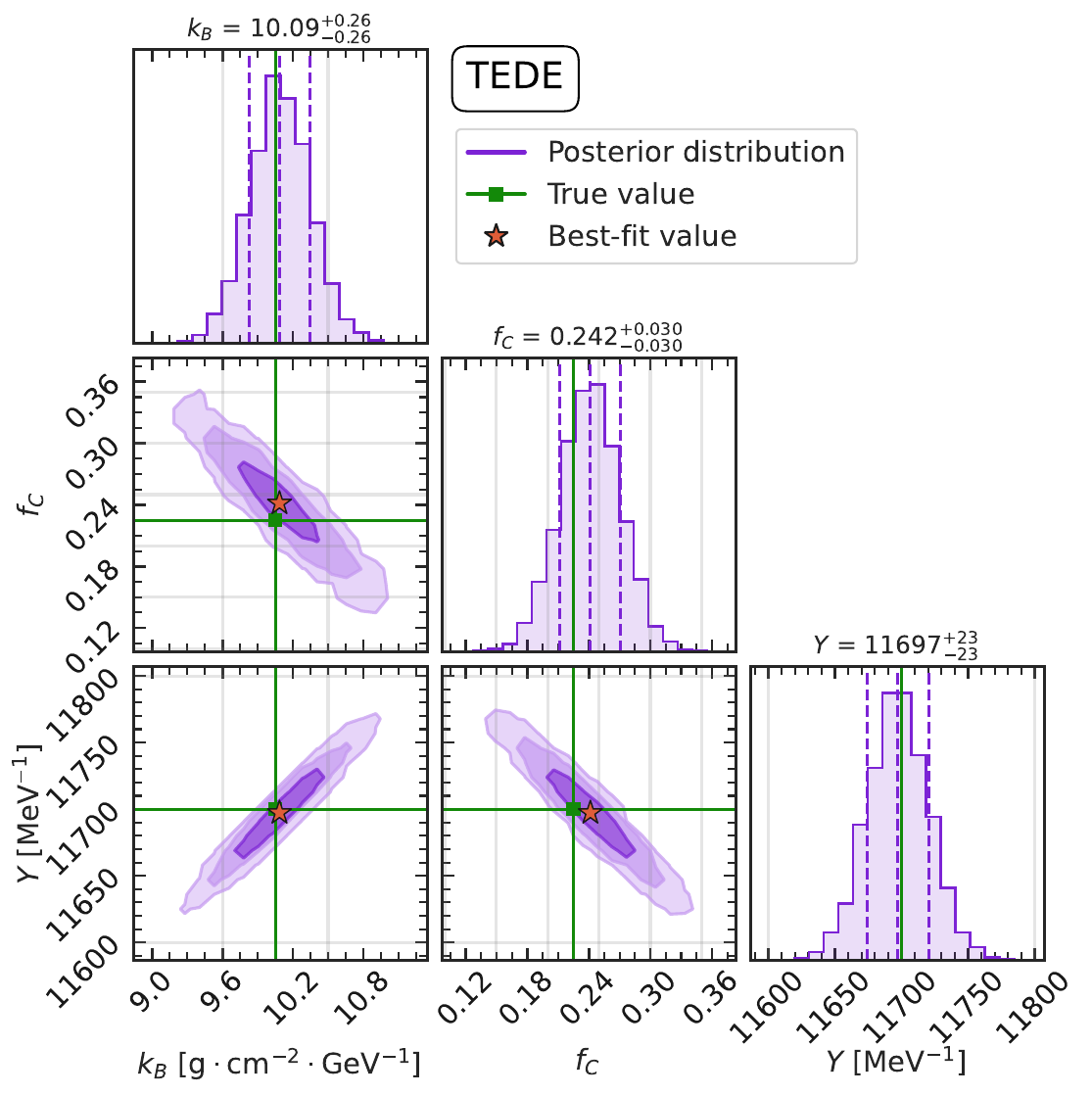}
  \includegraphics[width=0.49\textwidth]{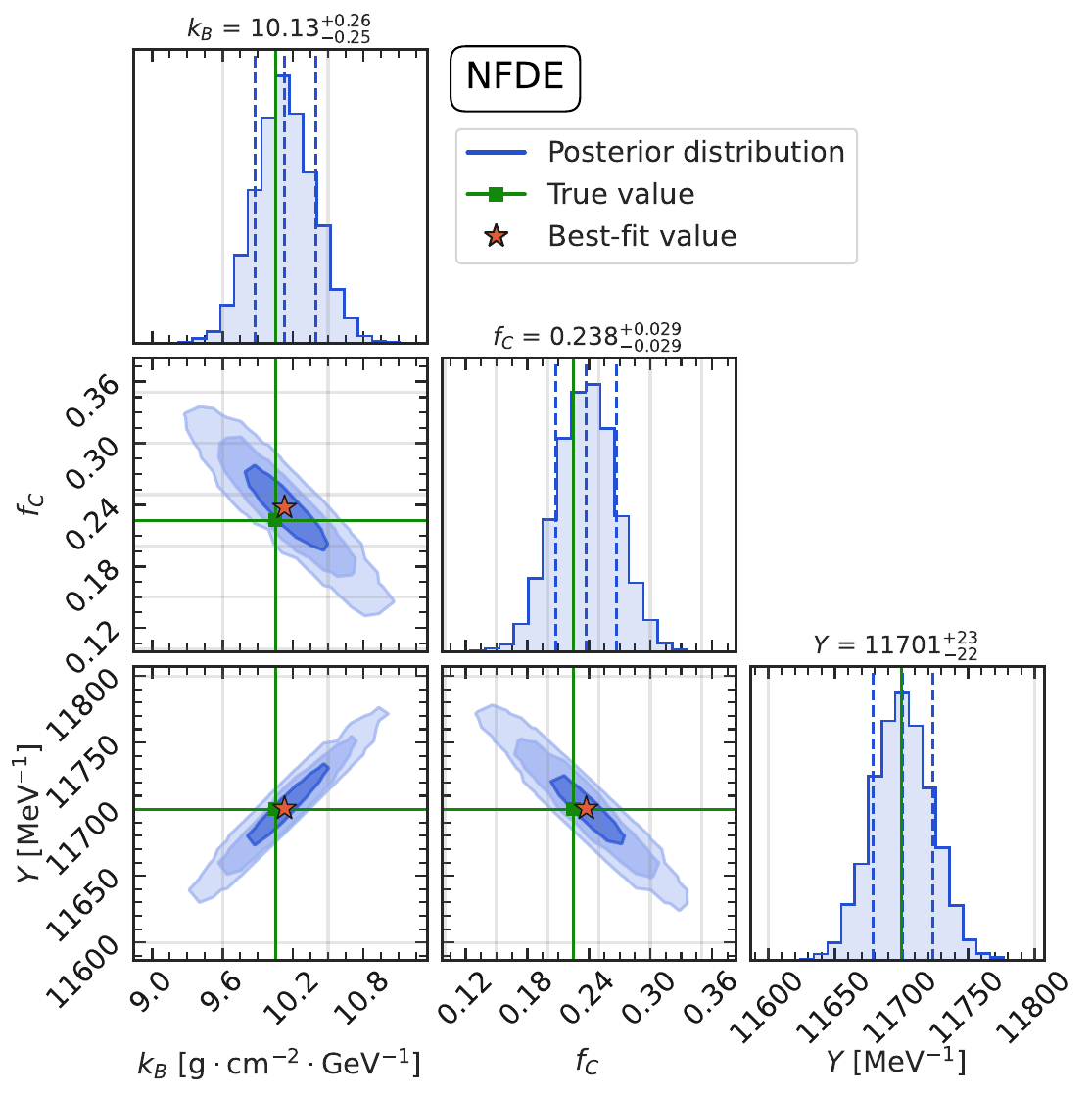}
  \caption{
      \textbf{Parameter estimation performance for a single
      representative point of testing dataset 1.} Corner plots show the one- and
      two-dimensional marginalized posterior probability distributions for the
      energy response parameters ($k_B, f_C, Y$), obtained using the TEDE
      (left) and NFDE (right) models with the Bayesian nested sampling algorithm
      (from \texttt{ultranest}). Green lines indicate the true parameter values,
      while orange stars show the best-fit values. The gray grid represents the
      parameter values of the training dataset. Both models yield
      posterior distributions centered near the true values, demonstrating
      the effective parameter recovery through the
      combined fit across five calibration sources.
  }
  \label{fig:fit_tede_and_nfde}
\end{figure*}

To demonstrate the parameter recovery capabilities of the models,
we first examine a representative example from testing dataset 1, corresponding
to true parameters $k_B=10.05~\text{ g} \cdot \text{cm}^{-2}  \cdot \text{GeV}^{-1}$, $f_C =
0.225$, $Y = 11{,}700 \text{ MeV}^{-1}$. As detailed in~\autoref{sec:methods}, for each point and each source of testing dataset 1, we generate 10,000 events. Due to the decay scheme of the neutron sources, where both a neutron and de-excitation gammas are emitted, the final amount of events is doubled for these sources. Additionally, events with total charge collected below the trigger threshold (of around 200 photo-electrons) are discarded in the simulation. This effect varies by source: from several percent event loss for $^{137}$Cs to almost no event loss for $^{60}$Co, with the total event loss for all five sources combined being about 2.1\%. Thus, the total statistics for each point of testing dataset 1 is variable but approximately equal to 70,000 events. For simplicity, we define the term ``statistical exposure level'' as the number of generated events per source (i.e., 10,000 for testing dataset 1), which is therefore approximately seven times smaller than the actual total statistics of the combined sample.

A joint fit incorporating the predicted energy spectra for all five calibration sources was performed using likelihood functions for statistical inference that are attributed to each model's approach as described in detail in~\autoref{subsec:parameter_estimation}. These are a binned Poisson log-likelihood for TEDE, which naturally accommodates its histogram-based
representation, and an extended unbinned likelihood~\cite{extended_ll_cowan, extended_ll_barlow} for
NFDE, which benefits from its ability to evaluate exact probability densities at individual data
points. In addition to the three energy response parameters ($k_B$, $f_C$, $Y$), the fit simultaneously determines the normalization constants $N_{s=1\dots5}$ for each of the five calibration sources: ${}^{137}$Cs, ${}^{40}$K, ${}^{60}$Co, ${}^{241}$Am-Be, and ${}^{241}$Am-${}^{13}$C, respectively. These $N_s$ parameters represent the expected total number of events and are treated as free parameters.

The Bayesian nested sampling algorithm~\cite{buchner2023nested}
(from \texttt{ultranest}~\cite{ultranest}) was employed to infer the energy response parameters for
both models. The posterior distributions shown in  \autoref{fig:fit_tede_and_nfde} reveal
well-constrained parameters, with estimates that are both precise and accurate. Using multiple
calibration sources helps effectively reduce the uncertainties in energy response modeling. Both
models achieve comparable precision, successfully recovering the input parameters within their
uncertainties, as summarized in the Table~\ref{tab:parameter_comparison}.

\begin{table}[!htb]
\centering
\setlength{\tabcolsep}{4pt}
\begin{tabular}{|l|c|cc|}
\hline
\multirow{2}{*}{Parameter} & \multirow{2}{*}{True value} & \multicolumn{2}{c|}{Estimated value} \\
\cline{3-4}
& & TEDE & NFDE \\
\hline
\rule{0pt}{2.75ex}
$k_B$, $\text{g} \cdot \text{cm}^{-2} \cdot \text{GeV}^{-1}$ & $10.05$ & $10.09^{+0.26}_{-0.26}$ & $10.13^{+0.26}_{-0.25}$ \\[0.75ex]
$f_C$ & $0.225$ & $0.242^{+0.030}_{-0.030}$ & $0.238^{+0.029}_{-0.029}$ \\[0.75ex]
$Y$, $\text{MeV}^{-1}$ & $11{,}700$ & $11{,}697^{+23}_{-23}$ & $11{,}701^{+23}_{-22}$ \\[0.75ex]
\hline
\end{tabular}
\caption{\textbf{Parameter estimation results for a single representative point from testing dataset 1.} The table presents a comparison of true values of the energy response parameters and best-fit estimates (with their corresponding uncertainties, computed as the $34.1\%$ of the probability on either side of the median). Parameter inference is performed using the Bayesian nested sampling method from \texttt{ultranest} for both TEDE and NFDE models.}
\label{tab:parameter_comparison}
\end{table}

Notably, the distance between the training and the testing
grids demonstrates the models' ability to interpolate effectively between grid
points, a critical feature for practical applications where the true parameters
may lie anywhere within the parameter space.

\begin{figure}[!htb]
    \centering
    \includegraphics[width=0.49\textwidth]{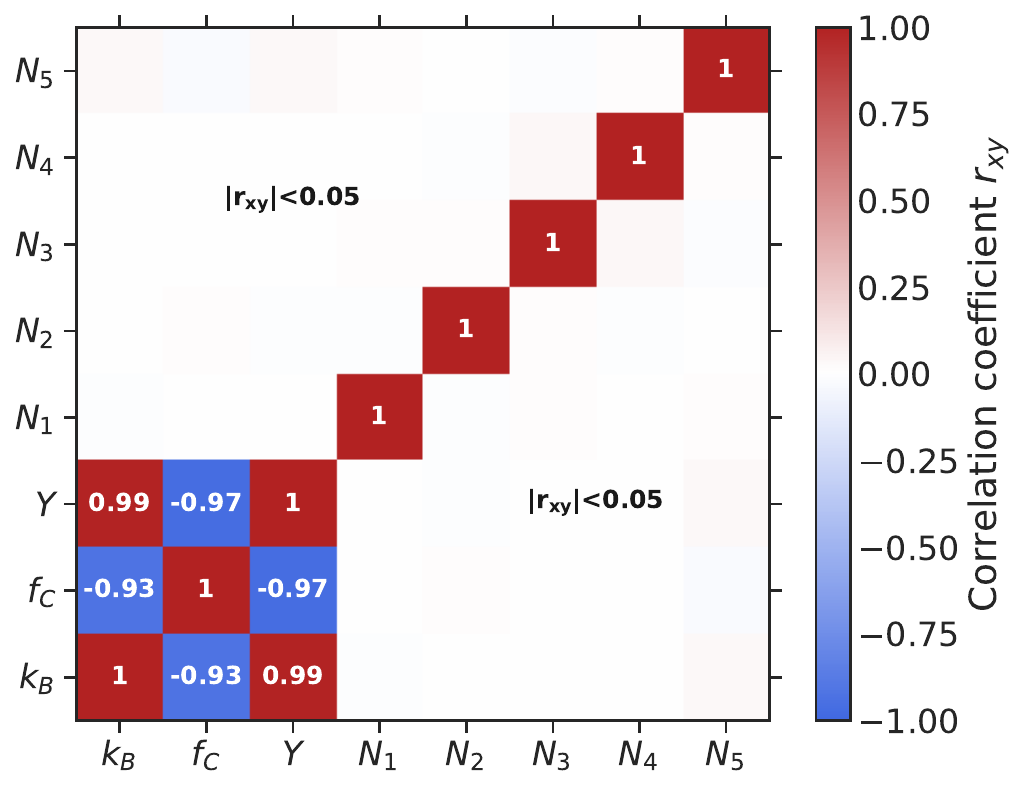}
    \caption{
      \textbf{Correlation coefficients $r_{xy}$ for the fit parameters.}
        The NFDE parameter estimation results from
        \autoref{fig:fit_tede_and_nfde} are used. The figure illustrates strong
        correlations and anti-correlations between the energy response parameters $k_B$,
        $f_C$, and $Y$. The normalization constants $N_s$ for each calibration source exhibit negligible
        correlations with other parameters. Here, $N_{s=1\dots5}$ correspond to
        $^{137}$Cs, $^{40}$K, $^{60}$Co, $^{241}$Am-Be, and $^{241}$Am-$^{13}$C, respectively.
    }
    \label{fig:corrcoeffs}
\end{figure}

The posterior distribution plots also reveal the correlations between
parameters which are quantitatively presented in \autoref{fig:corrcoeffs} for
the NFDE fit. Such high-correlation behavior is expected given parameters joint
influence on the non-linear effects of the energy response. The Bayesian nested sampling approach
effectively captures these correlations, providing a complete picture of
the parameter space including degeneracies and constraints.

\subsubsection{Aggregated results and the robustness analysis}
To assess the robustness and potential biases of our approach, we conducted a
comprehensive analysis across varying statistical conditions and throughout the parameter space.

\paragraph{Stability under varying statistical exposure levels.}

\begin{figure*}[!htb]
  \centering
      \includegraphics[width=1\textwidth]{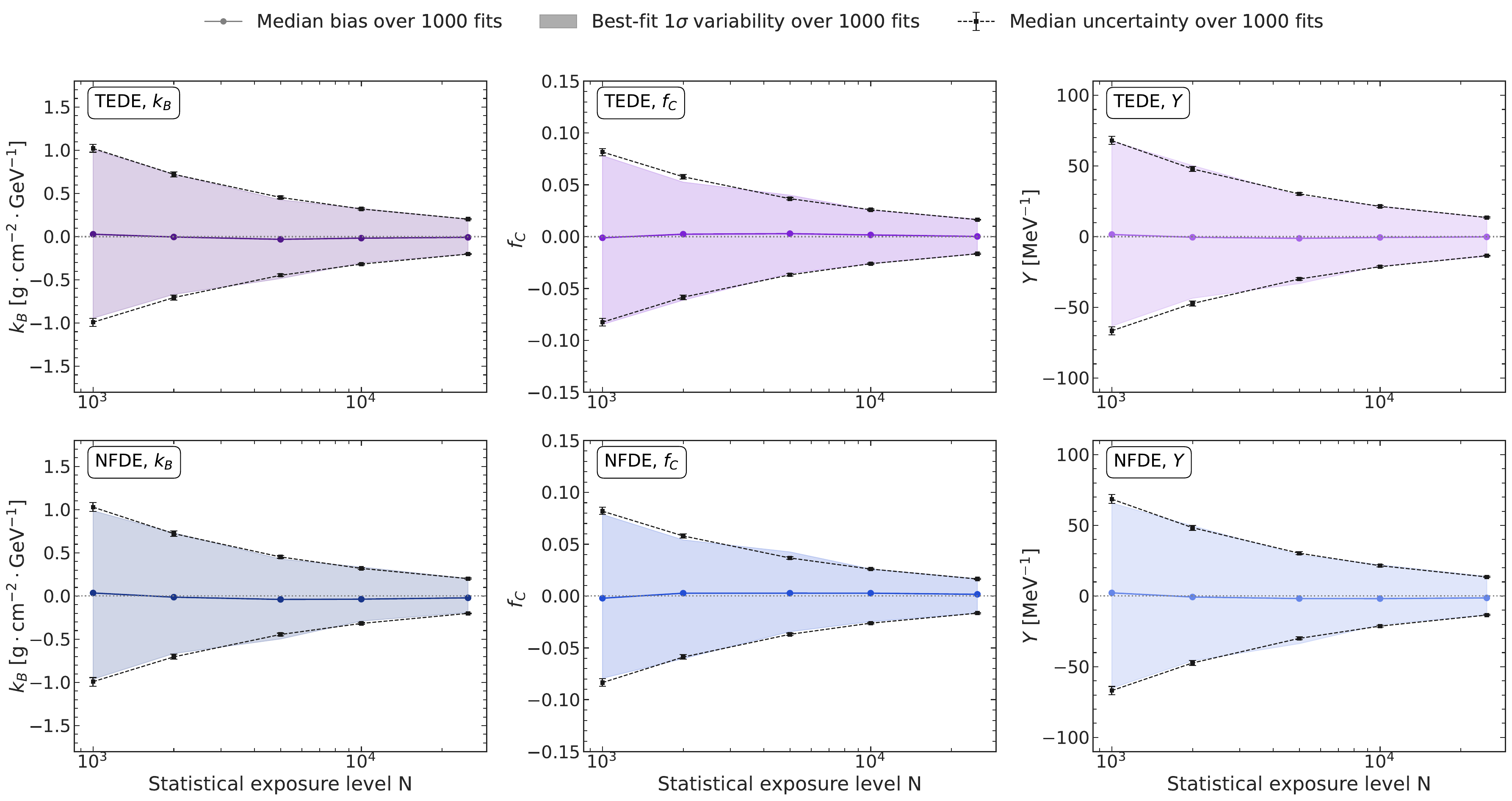}
  \caption{
    \textbf{Aggregated parameter estimation results using TEDE (top row) and NFDE (bottom row) models for testing dataset 2.} The plots show aggregated results for $k_B$ (left column), $f_C$ (middle column), and $Y$ (right column), including the median bias, the $1\sigma$ widths of the best-fit value distributions, and the estimated uncertainties. Solid lines with circle markers represent the median bias (i.e., the median of the distribution of differences between best-fit and true parameter values). Shaded regions indicate the $1\sigma$ widths of the best-fit value distributions (i.e., 68.3\% coverage of the obtained best-fit values). Dashed lines with square markers represent the medians of the $1\sigma$ uncertainty widths estimated by the fits. Finally, the black error bars show the $1\sigma$ widths of these estimated uncertainty distributions. Each point corresponds to results from 1,000 fits at a given exposure level. These results demonstrate the robustness and accuracy of both models across varying statistical scenarios, showing near-zero biases and estimated uncertainties that closely align with the actual variability in best-fit values.
}
\label{fig:results_td2}
\end{figure*}

The analysis based on testing data 2 corresponds to results obtained for a single point in the
($k_B, f_C, Y$) parameter space with varying statistical exposure levels. This point
is set at $k_B = 15.45 \text{ g} \cdot \text{cm}^{-2} \cdot \text{GeV}^{-1}$, $f_C = 0.525$, and $Y = 10{,}100 \text{ MeV}^{-1}$, positioned near the center of the parameter space and midway between training grid points.

\autoref{fig:results_td2} presents a comparative assessment of the models,
TEDE and NFDE, studying their performance at a single point in the parameter
space across varying statistical exposure levels. Each data point corresponds
to 1,000 individual fits, with a total of 10,000 fits conducted
across five exposures and two models.
One key observation is that both models exhibit a near-zero median bias (colored solid lines), with relative deviations below 0.25\% for $k_B$, 0.55\% for $f_C$, and 0.02\% for $Y$, calculated with respect to their true values. In addition, the medians of the estimated uncertainties (black solid lines) consistently align with the actual 1$\sigma$ widths of the distributions of best-fit values (shaded regions).
Here, the 1$\sigma$ widths represent intervals that enclose 34.1\% of the probability on either side of the median in the best-fit distributions. The reported medians of the estimated uncertainties, along with the medians of the lower and upper 1$\sigma$ bounds, are derived from 1,000 individual fits. These findings indicate that neither model systematically skews parameter estimations, reinforcing confidence in the overall accuracy of the best-fit estimates. Additionally, the fit results for TEDE and NFDE (black lines on~\autoref{fig:results_td2}) show negligible differences in the uncertainty estimates, highlighting alignment of the results not only within each model but also between them.

Moreover, for both models, the estimated uncertainties closely follow the expected dependence on the number of events per source, as dictated by pure statistical errors~\cite{Cowan:1998ji}. Specifically, the uncertainties scale as $1/\sqrt{N}$, where $N$ is the statistical exposure level. Thus, the proportionality constants are:
\begin{gather*}
\frac{0.32\ \text{g} \cdot \text{cm}^{-2} \cdot \text{GeV}^{-1}}{\sqrt{N/10^4}}, \quad
\frac{0.026}{\sqrt{N/10^4}}, \quad
\frac{21\ \text{MeV}^{-1}}{\sqrt{N/10^4}},
\end{gather*}
for $k_B$, $f_C$, and $Y$, respectively, with the constants extracted by fitting the squared median uncertainties as a linear function of $1/N$. At the benchmark exposure level of $10^4$, this translates to relative uncertainties of approximately $2.1\%$ for $k_B$, $4.9\%$ for $f_C$, and $0.2\%$ for $Y$.

Finally, the $1\sigma$ widths of the estimated uncertainty distributions (error bars of the corresponding points of black solid lines)  are rather small for both models, indicating fit stability and reliability of obtained best-fit values and their estimated errors.

These findings demonstrate that both TEDE and NFDE models are robust, exhibiting negligible
systematic bias and producing uncertainty estimates that align well with expected statistical
uncertainty. Their accuracy remains consistent across different exposure levels, and uncertainties
scale precisely with the number of events per source, confirming their reliability for statistical analyses.

\paragraph{Stability across varying parameter space.}

\begin{figure*}[!htb]
  \centering
  \includegraphics[width=1\textwidth]{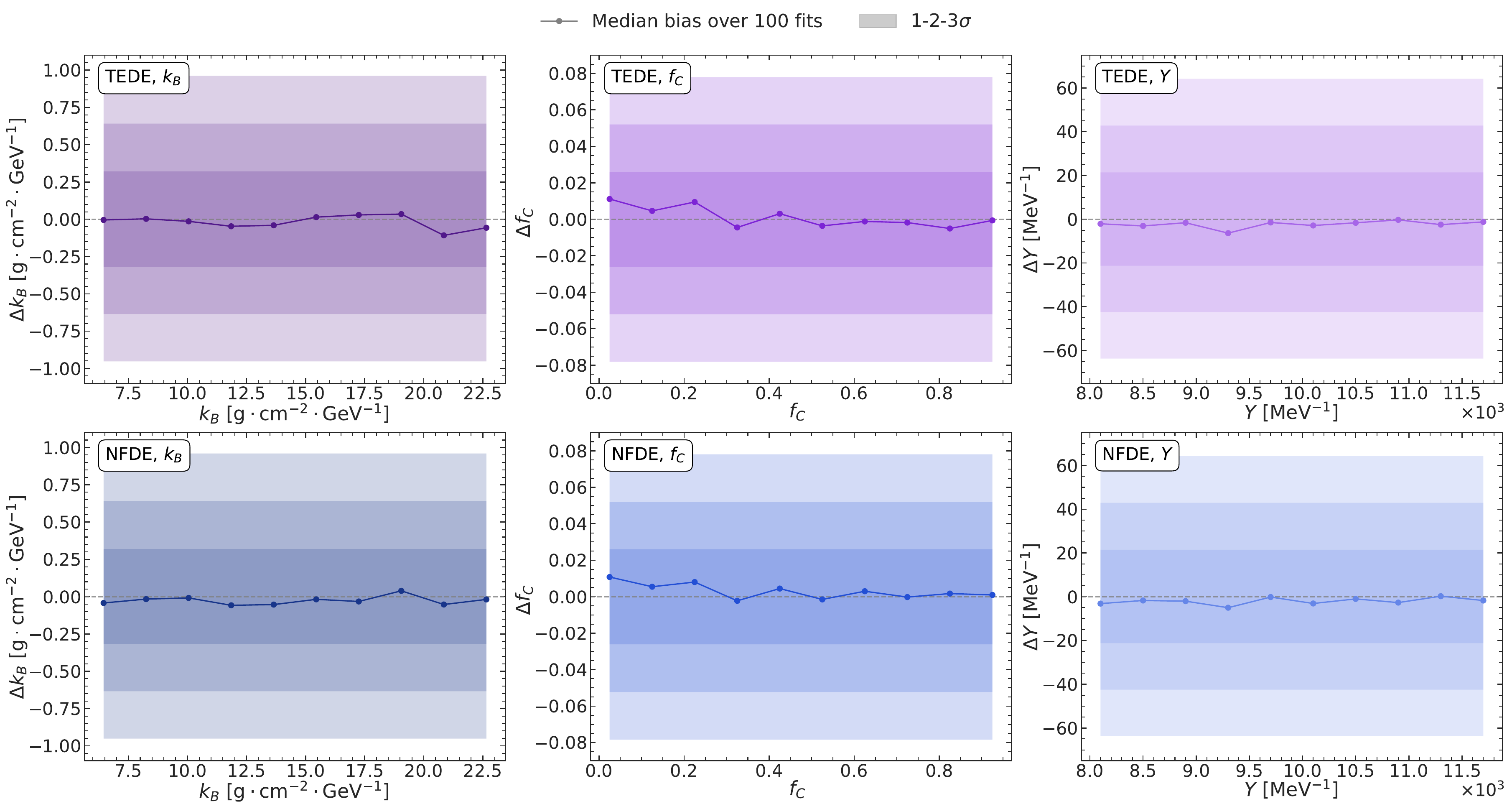}
  \caption{
      \textbf{Performance across the considered parameter space using
      TEDE  (top row) and NFDE (bottom row), obtained for testing
      dataset 1.} The dataset comprises 1000 points sampling the ($k_B, f_C,
      Y$) space, each with 10,000 exposure level. Shaded areas represent 1,
      2, and 3$\sigma$ uncertainties obtained from testing dataset 2 for the
      same exposure level. Each point represents the median bias, defined as
      the difference between the best-fit and true parameter value, computed
      over 100 fits. Both models maintain near-zero bias  across the parameter space, demonstrating their robustness and stability.}
  \label{fig:results_td1}
\end{figure*}

This analysis is based on testing dataset 1 and investigates the behavior of the TEDE and NFDE
models under a fixed statistical exposure (10,000 events per source) while systematically varying
the parameter space. Each data point in \autoref{fig:results_td1} represents the aggregated results
of 100 individual fits, allowing for a detailed examination of bias behavior across the parameter
range. The number 100 reflects our parameter grid structure: when analyzing the
bias for a specific parameter value (e.g., one value of $k_B$), we aggregate results from 100 fits
corresponding to all combinations of the other two parameters (10 values of $f_C$ $\times$ 10 values
of $Y$) at that fixed $k_B$.

Both TEDE and NFDE models demonstrate biases that consistently lie within the 1$\sigma$ best-fit
width, indicating that neither model exhibits significant systematic deviations. This suggests
that the models are capable of successfully interpolating between points in the training grid,
maintaining accuracy and robustness across the parameter space. Furthermore, the observed bias
behavior reinforces confidence in their predictive reliability, ensuring stable performance even
under varying parameter conditions.

\subsubsection{Differences between the parameter minimization methods}

As discussed in~\autoref{subsec:parameter_estimation}, we employed three
different methods for parameter estimation: the Frequentist MIGRAD optimization
algorithm (from \texttt{iminuit}~\cite{iminuit}) and its MINOS uncertainty estimation,
alongside two Bayesian approaches (nested sampling from \texttt{ultranest} and a
self-written Metropolis-Hastings~\cite{metropolis, metropolis_full} Markov chain Monte Carlo (MCMC) method).

The analysis demonstrates the reliability of both TEDE and NFDE models across different parameter
estimation methods. Frequentist approaches (MIGRAD+MINOS) and Bayesian techniques (nested sampling
and Metropolis-Hastings MCMC) produce consistent best-fit values and confidence or credible
intervals, ensuring robustness in parameter inference.
While MIGRAD+MINOS excels in speed, making it ideal for rapid estimations, Bayesian methods provide a more comprehensive statistical understanding by delivering full posterior distributions and offering deeper insights into parameter behavior. The results obtained with MIGRAD+MINOS and Metropolis-Hastings MCMC can be found in~\autoref{ax:param_estimation_other_methods}.

This balance between efficiency and depth highlights the strengths
of each approach, underlying their applicability in diverse analysis scenarios.

\section{Discussion}
\label{sec:discussion}

NFDE's ability to produce continuous PDFs for unbinned analysis provides a potential advantage in statistical precision for parameter recovery, particularly in low statistics cases. Although in our comparative analyses this difference is negligible, it could be more significant in other applications. Nevertheless, TEDE offers computational efficiency and implementation simplicity while maintaining the same performance, serving as a complementary approach that validates the overall methodology. We would also like to highlight that computational considerations do not limit the adoption of unbinned likelihood methods nowadays. While traditionally viewed as computationally expensive, our implementation of the unbinned analyses can be efficiently parallelized on both GPU~\cite{andrea_gpu} and CPU with up to $M$ parallel units (where $M$ is the number of events). The increased availability of modern computing resources addresses a long-standing barrier to wider adoption of unbinned techniques in large-scale experiments. The scalability of these approaches is also promising when considering extensions to higher-dimensional parameter spaces or larger data volumes. Our current implementation of NFDE scales linearly with data size, while the parameter space complexity primarily affects the training phase rather than inference.

To provide quantitative context for the computational advantages of our approach, we present timing comparisons for the key components of parameter estimation. Our TEDE model takes approximately 3~ms to build each PDF, independently of the number of events in the dataset. 
The NFDE model requires approximately $40 + 0.002 \cdot M$ ms for probability density estimations, corresponding to 60 ms for a typical dataset of $10^4$ events. In contrast, the full event simulation and reconstruction, using the official JUNO software, takes approximately 15 hours per parameter point per CPU for $10^4$ events. Critically, this number of events is insufficient to construct reliable PDFs with negligible statistical fluctuations, necessitating orders of magnitude more events and correspondingly longer computation times. This represents a computational speedup of several orders of magnitude (in $\sim$$10^6$ times in the case of $10^4$ events), enabling previously intractable analyses such as comprehensive uncertainty quantification.

Despite using neural networks, both approaches maintain physical interpretability --- an important advantage over typical ``black box'' machine learning models~\cite{belle2021principles}. The explicit connection to the underlying physical parameters ($k_B$, $f_C$, $Y$) ensures that the results remain physically meaningful and that uncertainties directly translate to constraints on detector properties. 

To our knowledge, this work also represents a relatively rare successful application of one-dimensional conditional normalizing flows in an experimental physics context. While normalizing flows have gained significant attention in particle physics for tasks such as event generation, anomaly detection, and variational inference in high-dimensional settings~\cite{hepmllivingreview}, their use for one-dimensional conditional density estimation in particle physics has been limited.

We also demonstrate that the presented approach is capable of resolving highly correlated parameters. The energy response parameters in our study exhibit strong correlations (as shown in~\autoref{fig:corrcoeffs}), presenting a challenging parameter estimation scenario. Both models successfully explore this correlation structure.

Looking forward, several avenues for improvement exist. Higher-dimensional normalizing flows could potentially model the joint distribution of energy and other observables (such as time distribution or particle identification discriminators). Although directly extending to higher dimensions can be challenging, methods such as those presented in Ref.~\cite{dirmeiersimulation} have a potential to effectively address this issue. Prior constraints could also be incorporated into the Bayesian framework to further enhance parameter estimation when prior knowledge is available and justified to be used~\cite{Gelman:2013}. Extensions to time-dependent calibration, where detector parameters evolve, represent another promising direction for future research.

\section{Methods}
\label{sec:methods}

\subsection{Data description}
\label{subsec:data_description}

The well-established approach in the field of machine learning is to prepare three different categories of datasets: so-called training, validation, and testing datasets. The training dataset is used to train a model (to adjust the model's learnable weights by minimizing a loss function). The validation dataset is used to evaluate model performance \textit{during} training process and guides hyperparameter optimization, while the testing dataset provides a final unbiased quantification of model performance once it \textit{is trained}. These datasets must remain independent from one another to prevent biases and overfitting.

The data used in this study were generated using the official JUNO software~\cite{JUNO:juno_soft}, based on a detailed \texttt{Geant4}~\cite{geant4_1, geant4_2} detector simulation. Different processes during photo-electron collection and further current amplification (e.g., dark noise), as well as intrinsic effects of electronics, are also simulated. The waveform reconstruction is performed using the deconvolution method implemented in the official JUNO software based on Ref.~\cite{Huang:2017abb}. Our training, validation, and testing datasets are constructed to balance comprehensive coverage of the parameter space with the density of the points. We simulate three gamma sources (${}^{137}$Cs, ${}^{40}$K, ${}^{60}$Co) and two neutron sources (${}^{241}$Am-Be, ${}^{241}$Am-${}^{13}$C) placed at the detector center.

\paragraph{Training dataset.} To properly cover the region of interest of the energy response parameters space, we define a ($k_B$, $f_C$, $Y$) grid with 21 uniformly spaced values per each parameter as follows:
\[
\begin{aligned}
&\Theta^{\rm Train}_{k_B} = \{6 + 0.9j\} \ \text{ g} \cdot \text{cm}^{-2} \cdot \text{GeV}^{-1}, \\
&\Theta^{\rm Train}_{f_C} = \{0 + 0.05j\}, \\
&\Theta^{\rm Train}_Y = \{8000 + 200j\} \ \text{ MeV}^{-1}, \\
&\text{where } j\colon j \in \mathbb{Z},\, 0 \le j \le 20.
\end{aligned}
\]
The full training parameter space is then defined as $\Theta^{\rm Train} = \Theta^{\rm Train}_{k_B} \times \Theta^{\rm Train}_{f_C} \times \Theta^{\rm Train}_{Y}$, resulting in $21^3 = 9261$ unique parameter combinations. For each parameter point and each of the five calibration sources, we simulate $10^4$ events. Events with light collection below the trigger threshold ($\sim$200 photo-electrons) are discarded. For the neutron sources (${}^{241}$Am-Be, ${}^{241}$Am-${}^{13}$C), simulations yield prompt (de-excitation gamma and neutron recoil) and delayed (gamma from neutron capture) events, both of which are used together in a single distribution. Therefore, to construct datasets for models' training, $\sim$600 million events are simulated.

\paragraph{Validation datasets.} We produce two validation datasets for monitoring model performance during training and optimizing hyperparameters:

\textit{1) Validation dataset 1.} This dataset uses a $10^3 = 1000$ point grid, shifted relative to the training grid to lie exactly midway between training points, probing the models interpolation capability in an unbiased manner under the most difficult conditions:
\[
\begin{aligned}
&\Theta^{\rm Val_1}_{k_B} = \{7.35 + 1.8j\} \ \text{ g} \cdot \text{cm}^{-2} \cdot \text{GeV}^{-1}, \\
&\Theta^{\rm Val_1}_{f_C} = \{0.075 + 0.1j\}, \\
&\Theta^{\rm Val_1}_Y = \{8300 + 400j\} \ \text{ MeV}^{-1}, \\
&\text{where } j\colon j \in \mathbb{Z},\, 0 \le j \le 9.
\end{aligned}
\]
Thus, $\Theta^{\rm Val_1} = \Theta^{\rm Val_1}_{k_B} \times \Theta^{\rm Val_1}_{f_C} \times \Theta^{\rm Val_1}_{Y}$. Analogously to the training dataset, we simulate $10^4$ events per source per point, yielding $\sim$70 million events.

\textit{2) Validation dataset 2.} This dataset comprises three specific parameter points chosen from the validation dataset 1 grid space, but simulated with high statistics ($10^7$ events per source per point, $\sim$200 million total events):
\[
\begin{aligned}
\Theta^{\rm Val_2} = \{&(k_{B, i},\, f_{C, i},\, Y_i) \mid i = 1, 2, 3\} = \\
= \{&(\hspace{5pt} 7.35 \text{ g} \cdot \text{cm}^{-2} \cdot \text{GeV}^{-1},\, 0.075,\, \hspace{5pt} 8300 \text{ MeV}^{-1}), \\
&(14.55 \text{ g} \cdot \text{cm}^{-2} \cdot \text{GeV}^{-1},\, 0.475,\, \hspace{5pt} 9900 \text{ MeV}^{-1}), \\
&(23.55 \text{ g} \cdot \text{cm}^{-2} \cdot \text{GeV}^{-1},\, 0.975,\, 11900 \text{ MeV}^{-1})\}
\end{aligned}
\]
Although this dataset covers only three different points, the significantly higher statistics per spectrum (1000 times greater than validation dataset 1) provides statistically precise ground truth spectra, enabling accurate evaluation of model performance with minimal statistical uncertainties.

\paragraph{Testing datasets.} Two testing datasets, independent from the training and validation ones, are used for final model evaluation.

\textit{1) Testing dataset 1.} Similar to validation dataset 1, this uses a $10^3 = 1000$ point grid, again shifted to lie midway between training points, but different from the validation grid:
\[
\begin{aligned}
&\Theta^{\rm Test_1}_{k_B} = \{6.45 + 1.8j\} \ \text{ g} \cdot \text{cm}^{-2} \cdot \text{GeV}^{-1}, \\
&\Theta^{\rm Test_1}_{f_C} = \{0.025 + 0.1j\}, \\
&\Theta^{\rm Test_1}_Y = \{8100 + 400j\} \ \text{ MeV}^{-1}, \\
&\text{where } j\colon j \in \mathbb{Z},\, 0 \le j \le 9.
\end{aligned}
\]
Thus, $\Theta^{\rm Test_1} = \Theta^{\rm Test_1}_{k_B} \times \Theta^{\rm Test_1}_{f_C} \times \Theta^{\rm Test_1}_{Y}$. As with the other grid-based datasets, we simulate $10^4$ events per source per point ($\sim$70 million total). This dataset is used to assess the uniformity of parameter estimation performance across the entire parameter space, see~\autoref{fig:results_td1}.

\textit{2) Testing dataset 2.} This dataset focuses on a single parameter point near the center of the parameter space ($\{15.45 \text{ g} \cdot \text{cm}^{-2} \cdot \text{GeV}^{-1},\, 0.525,\, 10100 \text{ MeV}^{-1}\}$), also located midway between training points. For this point, we produce 1000 independent simulations (using different random seeds) for each of five different event exposures per source: $\{10^3, \ 2\times10^3, \ 5\times10^3, \ 10^4, \ 2.5\times10^4\}$ events. This extensive dataset ($\sim$300 million total events) allows us to study how parameter uncertainties and biases evolve as a function of event statistics. Moreover, it enables a comparison between the uncertainty estimations derived from individual fits and the actual distribution of best-fit values, therefore validating the uncertainty estimation procedure and evaluating potential systematic effects, see~\autoref{fig:results_td2}.

\begin{figure}[!t]
    \centering
    \includegraphics[width=0.49\textwidth]{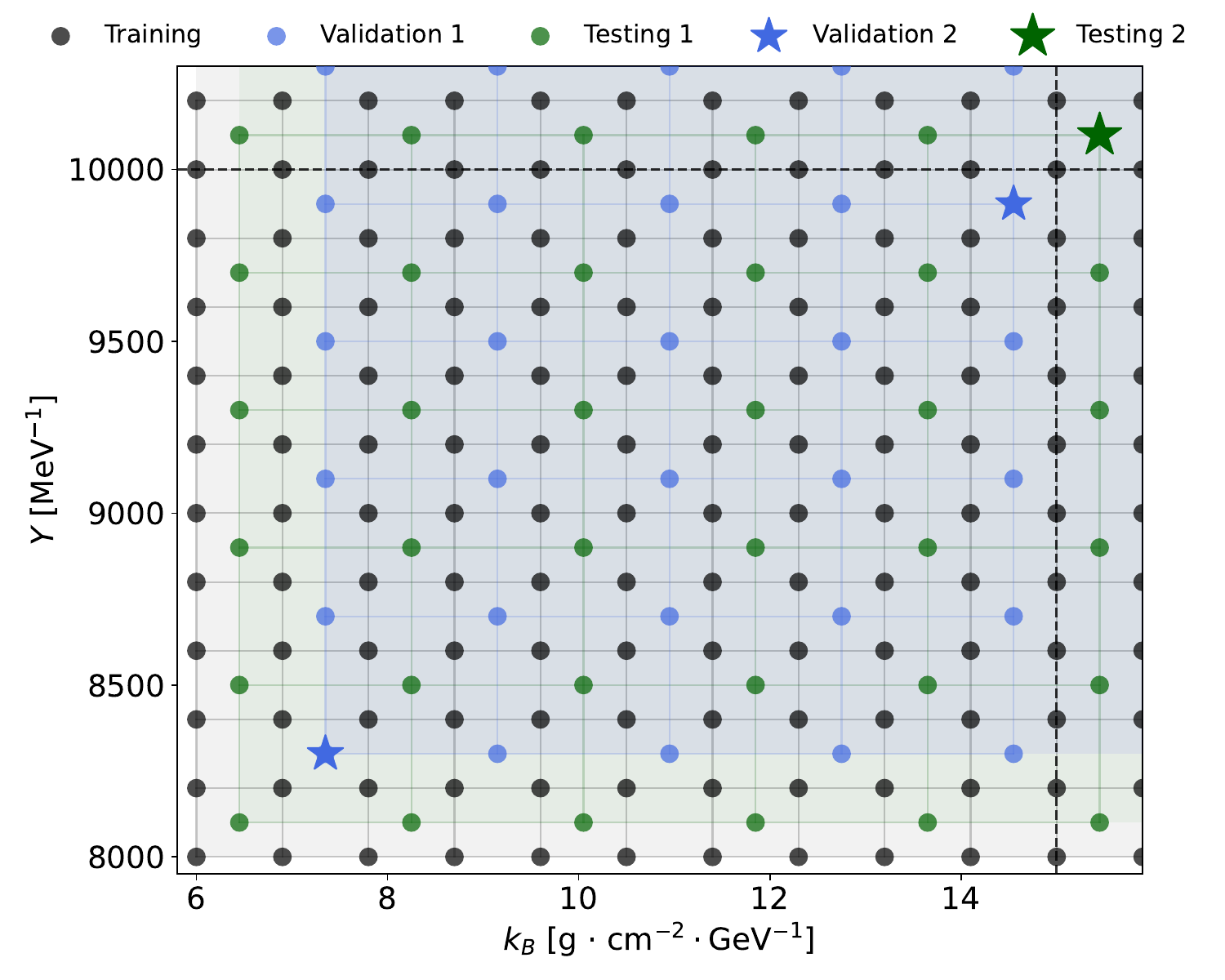}
    \caption{\textbf{Visualization of the parameter grid structure.} The figure shows a two-dimensional $k_B$-$Y$ projection of the three-dimensional parameter space. Training points (black) form a dense grid with step sizes of $0.9 \text{ g} \cdot \text{cm}^{-2} \cdot \text{GeV}^{-1}$ for $k_B$ and 200 MeV$^{-1}$ for $Y$. The validation dataset 1 grid (blue) and the testing dataset 1 grid (green) are positioned to lie midway between training points, creating the most challenging interpolation task for the models. High‑statistics points from validation dataset 2 and testing dataset 2 are shown as star markers. For testing dataset 2, at each point in the parameter space, 1000 independent datasets at five different exposure levels are produced. The central dashed lines indicate $k_B = 15 \text{ g} \cdot \text{cm}^{-2} \cdot \text{GeV}^{-1}$ and $Y = 10000 \text{ MeV}^{-1}$. For clarity, only a portion of the full projection is shown, with the actual grid extending to $k_B = 24 \text{ g} \cdot \text{cm}^{-2} \cdot \text{GeV}^{-1}$ and $Y = 12000 \text{ MeV}^{-1}$.}
    \label{fig:parameter_grid_vis}
\end{figure}

The spatial relationship between these different parameter grids is shown in~\autoref{fig:parameter_grid_vis}, which presents a two-dimensional projection of the $k_B$-$Y$ parameter space. This structured approach to dataset design allows us to comprehensively evaluate model performance: training on a dense grid while testing on points that lie exactly midway between training points creates the most challenging conditions for interpolation, ensuring that our evaluation measures the models' ability to generalize to unseen regions of the parameter space rather than simply memorize training examples. The carefully arranged validation and testing grids, offset from each other and from the training grid, enable robust cross-validation and ensure unbiased performance assessment.

\subsection{Machine learning approach}
\label{subsec:ml_models}

Directly using the full chain of the JUNO software within the inference frameworks (\autoref{subsec:parameter_estimation}) for energy response parameter tuning is computationally prohibitive due to the extensive time required for producing sufficient event statistics across the parameter space. To overcome this bottleneck, we develop neural likelihood estimators designed to emulate the simulation chain. These models provide rapid estimates of the conditional probability density function (PDF) of the energy spectrum, $p(x | \bm{\psi})$, where $x$ represents the visible energy (in number of reconstructed photo-electrons $N_{p.e.}$) and $\bm{\psi} = \{k_B, f_C, Y, S\}$ is the vector of conditioning variables comprising the energy response parameters ($k_B, f_C, Y$) and the calibration source type ($S$). Once trained using the simulation dataset, these fast surrogate models replace the full simulation during the parameter estimation procedures.

Thus, the core task for the ML models is formulated as one-dimensional conditional density estimation. This task becomes challenging due to the complex and multi-modal nature of the data. We investigate two different approaches as powerful tools to achieve the goals: 1) a direct multi-output regressor and 2) a conditional normalizing flows model. While, both of our approaches aim to approximate the conditional densities, the regressor approach provides an estimation through representing the density function using a histogram, and the normalizing flows solution offers the exact likelihood modeling. These strategies require different model architectures, training details, and likelihood formulations for subsequent parameter estimation, as detailed below.

\subsubsection{Transformer Encoder Density Estimator (TEDE)}

\textit{a) The method.} This direct regression approach begins by approximating the one-dimensional target continuous PDF $p(x|\bm{\psi})$ with a histogram using a fixed binning scheme. Let this histogram-based approximation, derived from the JUNO software output for a given condition $\bm{\psi}$, be represented by the vector $\bm{p}^{\rm JSW} = \{p^{\rm JSW}_1, p^{\rm JSW}_2, \dots, p^{\rm JSW}_{N_b}\}$. Here, $\rm JSW$ stands for ``JUNO software''. Each element $p_i^{\rm JSW}$ corresponds to the probability density in the $i$-th bin, calculated as $p_i^{\rm JSW} = n_i / (N_{\rm tot} \cdot \Delta x)$, where $n_i$ is the event count in bin $i$, $N_{\rm tot}$ is the total number of events for that condition, and $\Delta x$ is the bin width. By definition, this target distribution is normalized such that its discrete integral equals one: $\sum_{i=1}^{N_b} p^{\rm JSW}_i \cdot \Delta x = 1$.
The TEDE model, denoted as $h_{\bm{\theta}}$ with learnable parameters $\bm{\theta}$, is then trained to learn the direct mapping from the input conditions $\bm{\psi}$ to the vector of estimated PDF values $\bm{\hat{p}}^{\rm TEDE} = \{\hat{p}^{\rm TEDE}_1, \hat{p}^{\rm TEDE}_2, ..., \hat{p}^{\rm TEDE}_{N_b}\}$ for the corresponding energy bins:
\begin{equation}
\bm{\hat{p}}^{\rm TEDE} = h_{\bm{\theta}}(\bm{\psi}),
\end{equation}
where $\sum_{i=1}^{N_b} \hat{p}^{\rm TEDE}_i \cdot \Delta x = 1$ must hold.

We discretize the $x$ space into $N_b = 800$ bins with the bin width of $\Delta x = 20 \ N_{p.e.}$, spanning from 400 to 16400 $N_{p.e.}$. The core of the model is a transformer encoder architecture~\cite{transformer}, which is widely used in the ML literature~\cite{transformer_review_ml_1, transformer_review_ml_2} and becoming popular in particle physics applications in many different domains as well~\cite{transformer_physics, transformer_dinamo, spinner2024lorentz, wang2024interpreting, Pata:2023rhh, Mokhtar:2025zqs, Dreyer:2024bhs}.

\begin{figure*}
    \centering
    \includegraphics[width=1\linewidth]{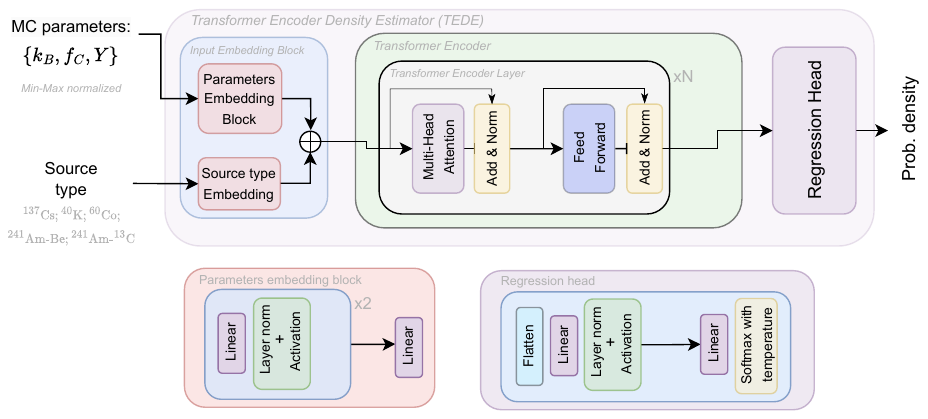}
    \caption{\textbf{Schematic representation of the TEDE model architecture.} The model takes min-max normalized energy response parameters ($k_B, f_C, Y$) and source type as inputs, processing them through dedicated embedding layers. The outputs of the embeddings are concatenated and fed into a transformer encoder comprising several encoder layers with multi-head attention. The transformer encoder's output is then flattened and processed by a regression head (with the softmax activation function as the final layer) that produces bin-by-bin probability density estimates for the corresponding energy spectrum. The transformer architecture is used as a powerful tool to capture complex relationships between the input energy response parameters and the calibration sources PDFs.}
    \label{fig:tede}
\end{figure*}

\textit{b) Architecture.} The TEDE model processes inputs through three main components as shown schematically in~\autoref{fig:tede}: (1) An input embedding block that transforms the input parameters $\bm{\psi}$ into high-dimensional vector representations. This block employs separate mechanisms for numerical and categorical inputs: the numerical MC parameters, after being min-max normalized to the range [0, 1], are processed using fully connected neural network with layer normalization~\cite{layer_norm}, while the categorical source type is embedded using a learnable lookup table. The resulting MC parameters and source type embeddings are then concatenated to form the input sequence for the next component, while each MC parameter is encoded with $n_{\rm tokens}$ --- a tunable hyperparameter. (2) A transformer encoder, the core of the model, consisting of several encoder layers (the exact number determined by hyperparameter optimization). Each layer utilizes multi-head self-attention mechanisms to capture complex dependencies among the embedded input features. Dropout regularization~\cite{dropout} can be applied within the encoder layers to mitigate overfitting; (3) A regression head, also constructed using a fully connected neural network with layer normalization, which processes the final representation from the transformer encoder. This head outputs a vector of $N_b=800$ raw values (logits), $\bm{z} = \{z_1, ..., z_{N_b}\}$. These logits are then passed through the final temperature-scaled softmax layer and divided by the bin width $\Delta x$ to create the output probability density estimates $\bm{\hat{p}}^{\rm TEDE}$: 
\begin{equation}
\hat{p}^{\rm TEDE}_i = \frac{1}{\Delta x} \frac{e^{z_i/T}}{\sum_{j=1}^{N_b} e^{z_j/T}},
\end{equation}
where $z_i$ are the logits (outputs of the linear layer preceding the softmax) for bin $i$, and $T$ is the softmax temperature hyperparameter. This construction inherently satisfies the PDF normalization requirement: $\sum_{i=1}^{N_b}\hat{p}_i^{\rm TEDE} \Delta x = 1$.  

\textit{c) Loss function.} As a standard information-theoretic measure for comparing probability distributions, we use the Kullback-Leibler (KL) divergence~\cite{kl_div, jordan1999introduction}, also known as relative entropy~\cite{dembo1991information}, that quantifies the discrepancy between the target $\bm{p}^{\rm JSW}$ and predicted $\bm{\hat{p}}^{\rm TEDE}$ distributions. We define the KL divergence from $\bm{\hat{p}}^{\rm TEDE}$ to $\bm{p}^{\rm JSW}$ as follows:
\begin{equation}
\begin{aligned}
D_{\rm KL}(\bm{p}^{\rm JSW} \parallel & \ \bm{\hat{p}}^{\rm TEDE}) = \sum_{i=1}^{N_b} p^{\rm JSW}_i \log \frac{p^{\rm JSW}_i}{\hat{p}^{\rm TEDE}_i} = \\
&= \sum_{i=1}^{N_b} p^{\rm JSW}_i (\log p^{\rm JSW}_i - \log \hat{p}^{\rm TEDE}_i),
\end{aligned}
\label{eq:kldiv}
\end{equation}
where the logarithm base is $e$. By convention, $0 \log 0 = 0$ and $\hat{p}^{\rm TEDE}_i > 0$ is always hold by construction of the softmax-based model's output.

A fundamental property of the KL divergence is established by Gibbs' inequality~\cite{gibbs_ineqv}, which states that for any probability distributions $P$ and $Q$:
\begin{equation}
D_{\rm KL}(P \parallel Q) \geq 0
\label{eq:gibbs}
\end{equation}
This inequality guarantees that the KL divergence is always non-negative. Furthermore, the equality $D_{\rm KL}(P \parallel Q) = 0$ holds if and only if the two distributions are identical. Gibbs' inequality provides the theoretical justification for using KL divergence as a loss function: minimizing $D_{\rm KL}(\bm{p}^{\rm JSW} \parallel \bm{\hat{p}}^{\rm TEDE})$ with respect to the TEDE parameters $\bm{\theta}$ directly encourages the model's output $\bm{\hat{p}}^{\rm TEDE}$ to approximate the true distribution $\bm{p}^{\rm JSW}$.

This direct mapping approach, based on histogram approximation of the density function employed by the TEDE model, serves as a benchmark for the exact conditional PDF modeling approach utilized by a normalizing flows model.

\subsubsection{Normalizing Flows Density Estimator (NFDE)}

\textit{a) The method.} This approach directly models the continuous conditional density $p(x|\bm{\psi})$ using the principle of normalizing flows~\cite{tabak_nf}. In out study, we consider conditional normalizing flows that learn an invertible and differentiable transformation $\bm{f}_{\bm{\theta}}$, parameterized by learnable parameters $\bm{\theta}$ and conditioned on $\bm{\psi}$. Since that main goal is conditional density estimation, we build inverted normalizing flows~\cite{trippe2018conditional} that maps the complex data distribution $p(x|\bm{\psi})$ to a simple base distribution $p_Z(z)$ (a standard normal $\mathcal{N}(0,1)$) for which the probability density is known analytically. The density $p(x|\bm{\psi})$ estimates are then computed via the change of variables formula:
\begin{equation}
\label{eq:nfde_change_of_variables}
\hat{p}(x|\bm{\psi}) = p_Z(\bm{f}_{\bm{\theta}}(x|\bm{\psi})) \left| \frac{\partial \bm{f}_{\bm{\theta}}(x|\bm{\psi})}{\partial x} \right|,
\end{equation}
where the Jacobian determinant term accounts for the change in volume induced by the transformation $\bm{f}_{\bm{\theta}}$. This approach allows for exact likelihood evaluation and efficient sampling from the learned distribution.

\begin{figure}
    \centering
    \includegraphics[width=0.49\textwidth]{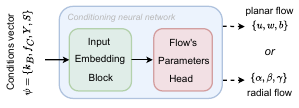}
    \caption{\textbf{Architecture of the conditioning neural network of the flows within the NFDE model.} This network enables the NFDE to learn conditional probability densities $p(x|\bm{\psi})$. For each flow in the sequence, a fully connected neural network takes the energy response MC parameters ($k_B, f_C, Y$) and the calibration source type ($S$) as input. The MC parameters, after being min-max normalized to the range [0, 1], are processed through a fully connected network, while the source type is converted into an embedding vector. These representations are concatenated and passed through a final fully connected network to generate the specific parameters (e.g., $w, u, b$ for a planar flow) required for that flow layer's transformation. This mechanism ensures that the overall transformation $\bm{f}_\theta$ applied by the NFDE is dependent on the conditioning variables, allowing the model to capture how the energy spectrum changes with different parameter settings and source types.}
    \label{fig:nfde_conditioning_network}
\end{figure}

\textit{b) Architecture.} Our NFDE implementation constructs the transformation $\bm{f}_{\bm{\theta}}$ as a sequence of $N_{\rm flows}$ flows. We consider two simple types of transformations: planar flows and radial flows~\cite{rezende2015variational}. The planar flow applies a simple, yet expressive, invertible transformation of the form $\bm{z}' = \bm{z} + u\tanh(w \bm{z} + b)$, where $\bm{z}$ is the input vector from the previous flow (or the original data $\bm{x}$ for the first flow), and $\bm{z}'$ is the output vector. This transformation provides local, direction-specific deformations of the input space. The radial flow applies a transformation of the form $\bm{z}' = \bm{z} + \frac{\alpha \beta (\bm{z} - \gamma)}{\alpha + |\bm{z} - \gamma|}$~\cite{trippe2018conditional}, which creates global, symmetric transformations around a reference point $\gamma$. 
These parameters ($\{w, u, b\}$ for planar flows; $\{\alpha, \beta, \gamma\}$ for radial flows) for \textit{each} flow are parametrized by a dedicated conditioning neural network, shown schematically in~\autoref{fig:nfde_conditioning_network}. Similarly to the TEDE's input processing presented on~\autoref{fig:tede}, this network takes the parameters $\bm{\psi} = \{k_B, f_C, Y, S\}$ as input and processes them through an input block. The numerical parameters ($k_B, f_C, Y$), after being min-max normalized to the range [0, 1], are processed by a fully connected network, and the categorical source type ($S$) is mapped to an embedding vector. These representations are concatenated and fed through another fully connected network to produce the required flow parameters. This conditioning mechanism makes the overall transformation $f_{\bm{\theta}}$ dependent on $\bm{\psi}$, enabling the model to learn how the energy spectrum's shape changes across different parameter settings and source types. By composing a sequence of such conditional flows, the model can capture highly complex, multi-modal conditional densities.

\textit{c) Loss function.} The NFDE model is trained by directly maximizing the likelihood of the training data, which is equivalent to minimizing the KL divergence between the model's likelihood and the target distribution~\cite{Papamakarios:2021}. For a dataset consisting of $N$ samples $\{x_i, \bm{\psi}_i\}$, the loss function is:
\begin{equation}
\begin{aligned}
&L(\bm{\theta}) = -\sum_{i=1}^{N} \log \hat{p}(x_i | \bm{\psi}_i) = \\
&=-\sum_{i=1}^{N} \left( \log p_Z(f_{\bm{\theta}}(x_i|\bm{\psi}_i)) + \log \left| \det \frac{\partial f_{\bm{\theta}}(x_i|\bm{\psi}_i)}{\partial x_i} \right| \right).
\label{eq:nfde_loss}
\end{aligned}
\end{equation}
This formulation allows the model to learn the full conditional PDF without requiring discretization (binning) or imposing strong parametric assumptions on the distribution's shape, leveraging the unbinned nature of the input data.

Both models were implemented in PyTorch~\cite{pytorch} and trained using PyTorch Lightning~\cite{pytorch_lightning} framework. During the training we also employ different distributed strategies --- GPU acceleration for TEDE and multi-CPU parallelization for NFDE --- to optimize use of available computational resources.

\subsection{Metrics to quantify models performance}
\label{subsec:metrics}

To systematically evaluate model performance during training and hyperparameter optimization, we employ a family of statistical distances based on the $L_p$-norm framework to quantify differences between probability distributions. This approach straightforwardly unifies three classical statistical distances --- Wasserstein, Cramér-von Mises, and Kolmogorov-Smirnov --- within a single mathematical formalism.
Specifically, for probability distributions $u$ and $v$ with respective cumulative distribution functions (CDFs) $U$ and $V$, the $L_p$-norm distance is defined as:
\begin{equation}
d_p(u, v) = \left( \int_{-\infty}^{+\infty} |U(x)-V(x)|^p \, dx \right)^{1/p}
\end{equation}
This formulation yields Wasserstein, Cramér-von Mises, and Kolmogorov-Smirnov distances depending on the choice of $p$:
\begin{itemize}
\item For $p = 1$, we obtain the Wasserstein distance, which measures the minimum ``work'' required to transform one distribution into another, making it particularly sensitive to shifts in distribution mean or location. Intuitively, it represents the minimum cost of transforming one distribution into another, where cost is proportional to the amount of probability mass moved multiplied by the distance it travels. It corresponds to the integrated absolute difference between the CDFs:
\begin{equation}
d_1(u, v) = \int_{-\infty}^{+\infty} |U(x)-V(x)| \, dx
\end{equation}
\item For $p = 2$, we obtain the Cramér-von Mises distance, which in the one-dimensional case is directly related to the energy distance (differing only by a factor of $\sqrt{2}$)~\cite{rizzo2016energy}. This metric balances sensitivity to both global distribution shifts and local deformations, effectively measuring the $L_2$-norm of the difference between CDFs:
\begin{equation}
d_2(u, v) = \sqrt{\int_{-\infty}^{+\infty} |U(x)-V(x)|^2 \, dx}
\end{equation}
\item For $p = \infty$, we obtain the Kolmogorov-Smirnov distance, which measures the maximum absolute difference between the two CDFs ($L_\infty$-norm) and is thus highly sensitive to localized discrepancies. This makes it particularly valuable for detecting localized anomalies that might be averaged out by other metrics:
\begin{equation}
d_\infty(u, v) = \sup_x|U(x)-V(x)|
\end{equation}
\end{itemize}
This unified $L_p$-norm framework enables consistent evaluation across both TEDE and NFDE approaches. While these metrics naturally accommodate the continuous probability densities of NFDE, the binned case of TEDE can be accomplished with the integral being approximated by a sum.
All three metrics were monitored throughout training and hyperparameter optimization, and we selected the Cramér-von Mises distance ($d_2$) as our primary optimization criterion due to its balanced sensitivity to both global distribution shifts and local deformations. This balance is particularly valuable when evaluating energy spectrum predictions across multiple calibration sources with diverse characteristics. We also found that $d_2$ distance provides the most robust translation to the ultimate goal of the models: energy response parameters estimation.

\subsection{Hyperparameter optimization and model training}
\label{subsec:hopt_and_training}

\begin{table*}[!htb]
\renewcommand*{\arraystretch}{1.25}
\centering
\begin{tabular}{|l@{\hskip 25pt}|l@{\hskip 25pt}|l@{\hskip 25pt}|}
\hline
\textbf{Hyperparameters} & \textbf{TEDE} & \textbf{NFDE} \\
\hline
\hline
\multicolumn{3}{|l|}{\textit{Model architecture hyperparameters}} \\
\hline
$n_{\rm tokens}$ & $\{k \mid k \in \mathbb{N}, 1 \leq k \leq 5\} \colon \bf{1}$ & \multicolumn{1}{c|}{--------} \\
$d_{\rm model}$ & $\{50k \mid k \in \mathbb{N}, 1 \leq k \leq 10\} \colon \bf{100}$ & \multicolumn{1}{c|}{--------} \\
$n_{\rm head}$ & $\{5, 10, 25\} \colon \bf{10}$ & \multicolumn{1}{c|}{--------} \\
$n_{\rm layers}$ & $\{k \mid k \in \mathbb{N}, 1 \leq k \leq 5\} \colon \bf{4}$ & \multicolumn{1}{c|}{--------} \\
$d_{\rm ff}$ & $\{32k \mid k \in \mathbb{N}, 1 \leq k \leq 15\} \colon \bf{384}$ & \multicolumn{1}{c|}{--------} \\
$n_{\rm flows}$ & \multicolumn{1}{c|}{--------} & $\{100 + 5k \mid k \in \mathbb{Z}, 0 \leq k \leq 20\} \colon \bf{120}$ \\
$n_{\rm units}$ & \multicolumn{1}{c|}{--------} & $\{10 + 5k \mid k \in \mathbb{Z}, 0 \leq k \leq 8\} \colon \bf{30}$ \\
Flow type & \multicolumn{1}{c|}{--------} & \textbf{\texttt{Planar}} (fixed) \\
Dropout rate & $\{p_{\rm{drop}} \mid 0.0 \leq p_{\rm{drop}} \leq 0.5 \} \colon \bf{0.0}$ & \multicolumn{1}{c|}{--------} \\
Activation function~\cite{relu, gelu, tanh, silu} & \texttt{ReLU}, \textbf{\texttt{GeLU}} & \texttt{ReLU}, \textbf{\texttt{GeLU}}, \texttt{Tanh}, \texttt{SiLU} \\
Softmax temperature & $\{T \mid 0.25 \leq T \leq 3.0 \} \colon \bf{2.03}$ & \multicolumn{1}{c|}{--------} \\
\hline
\hline
\multicolumn{3}{|l|}{\textit{General training hyperparameters}} \\
\hline
Optimizer~\cite{loshchilov2019decoupled, tieleman2012lecture} & \textbf{\texttt{AdamW}}, \texttt{RMSprop} & \textbf{\texttt{AdamW}} (fixed) \\
Learning rate & $\{\eta \mid 10^{-5} \leq \eta \leq 10^{-2}\} \colon \bf{1.9 \times 10^{-3}}$ & $\bf{10^{-4}}$ (fixed) \\
Scheduler type~\cite{exp, cosann} & 
\begin{tabular}[t]{@{}l@{}}
\vspace{-3pt}
\textbf{\texttt{Exponential}}, \texttt{ReduceOnPlateau}, \\
\texttt{CosineAnnealing}, \texttt{None}
\end{tabular} & \texttt{ReduceOnPlateau}, \textbf{\texttt{CosineAnnealing}} \\
Weight decay & $\{\lambda \mid 10^{-6} \leq \lambda \leq 10^{-1}\} \colon \bf{8 \times 10^{-5}}$ & $\bf{10^{-4}}$ (fixed) \\
Batch size & $\{2^{k} \mid k \in \mathbb{N}, 4 \leq k \leq 9\} \colon \bf{64}$ & $\bf{50}$ (fixed, 1 per a CPU unit) \\
\hline
\multicolumn{3}{|l|}{\textit{Optimizer-specific hyperparameters}} \\
\hline
\texttt{AdamW} decay rate $\beta_1$ & $\{\beta_1 \mid 0.5 \leq \beta_1 \leq 0.95\} \colon \bf{0.930}$ & $\{\beta_1 \mid 0.5 \leq \beta_1 \leq 0.95\} \colon \bf{0.87}$ \\
\texttt{AdamW} decay rate $\beta_2$ & $\{\beta_2 \mid 0.9 \leq \beta_2 \leq 0.999\} \colon \bf{0.929}$ & $\{\beta_2 \mid 0.9 \leq \beta_2 \leq 0.999\} \colon \bf{0.901}$ \\
\texttt{RMSprop} parameter $\alpha$ & $\{\alpha \mid 0.9 \leq \alpha \leq 0.999\} \colon \bf{-}$ & \multicolumn{1}{c|}{--------} \\
\hline
\multicolumn{3}{|l|}{\textit{Scheduler-specific hyperparameters}} \\
\hline
Cosine annealing period $T_{\rm max}$ & $\{T_{\rm max} \mid 5 \leq T_{\rm max} \leq 75\} \colon \bf{-}$ & $\{T_{\rm max} \mid 5 \leq T_{\rm max} \leq 75\} \colon \bf{50}$ \\
Exponential rate $\tau$ & $\{\tau \mid 0.8 \leq \tau \leq 0.99\} \colon \bf{0.92}$ & \multicolumn{1}{c|}{--------} \\
Reduce on plateau factor $\gamma$ & $\{\gamma \mid 0.8 \leq \gamma \leq 0.99\} \colon \bf{-}$ & $\{\gamma \mid 0.8 \leq \gamma \leq 0.99\} \colon \bf{-}$ \\
\hline
\hline
\multicolumn{3}{|l|}{\textit{Optimization procedure setting}} \\
\hline
Number of trials & \multicolumn{1}{c|}{\textbf{250}} & \multicolumn{1}{c|}{\textbf{25}} \\
Accelerator & \multicolumn{1}{c|}{GPU} & \multicolumn{1}{c|}{50 parallel CPUs} \\
\hline
\end{tabular}
\caption{\textbf{Comprehensive comparison of hyperparameter optimization for the TEDE and NFDE models.} The table presents search spaces and selected values (in \textbf{bold}) for both models, with rows organized into several subsections by hyperparameters application domain as follows: model architecture, general training, optimizer, learning rate scheduler, and the optimization procedure, respectively. a) For TEDE, the model architecture hyperparameters that define the model's structure are: $n_{\rm tokens}$ (the number of tokens that encode each input MC parameter), $d_{\rm model}$ (the embedding dimension size of the transformer encoder), $n_{\rm head}$ (the number of attention heads), $n_{\rm layers}$ (the number of encoder layers), and $d_{\rm ff}$ (the encoder's feed-forward dimension), dropout rate (controls the regularization strength), and the softmax temperature (the scaling factor for output probability distribution). b) For NFDE, the corresponding hyperparameters are: $n_{\rm flows}$ (the number of flows), $n_{\rm units}$ (the number of hidden units in each flow's conditioning neural network), and flow type (the class of the transformations; kept fixed as \texttt{Planar} based on preliminary experiments). c) Common parameters include activation functions, optimizer type, and scheduler type (with their specific hyperparameters). d) For optimizer hyperparameters, \texttt{AdamW} uses $\beta_1$ and $\beta_2$ as exponential decay rates for moment estimates, while \texttt{RMSprop} uses $\alpha$ as the smoothing constant. e) For schedulers, $\tau$ represents the decay rate for \texttt{Exponential} and $\gamma$ is the reduction factor for \texttt{ReduceOnPlateau}, while $T_{\rm max}$ is the half-cycle period for \texttt{CosineAnnealing}. f) Finally, the weight decay controls $L_2$ regularization strength, and batch size determines the number of different spectra processed simultaneously. Both models were optimized using Tree-structured Parzen Estimator, minimizing Cramér-von Mises distance ($d_2$) on the validation datasets.}
\label{tab:hyperparameters_comparison}
\end{table*}

\begin{figure*}[!t]
    \centering
    \includegraphics[width=1\linewidth]{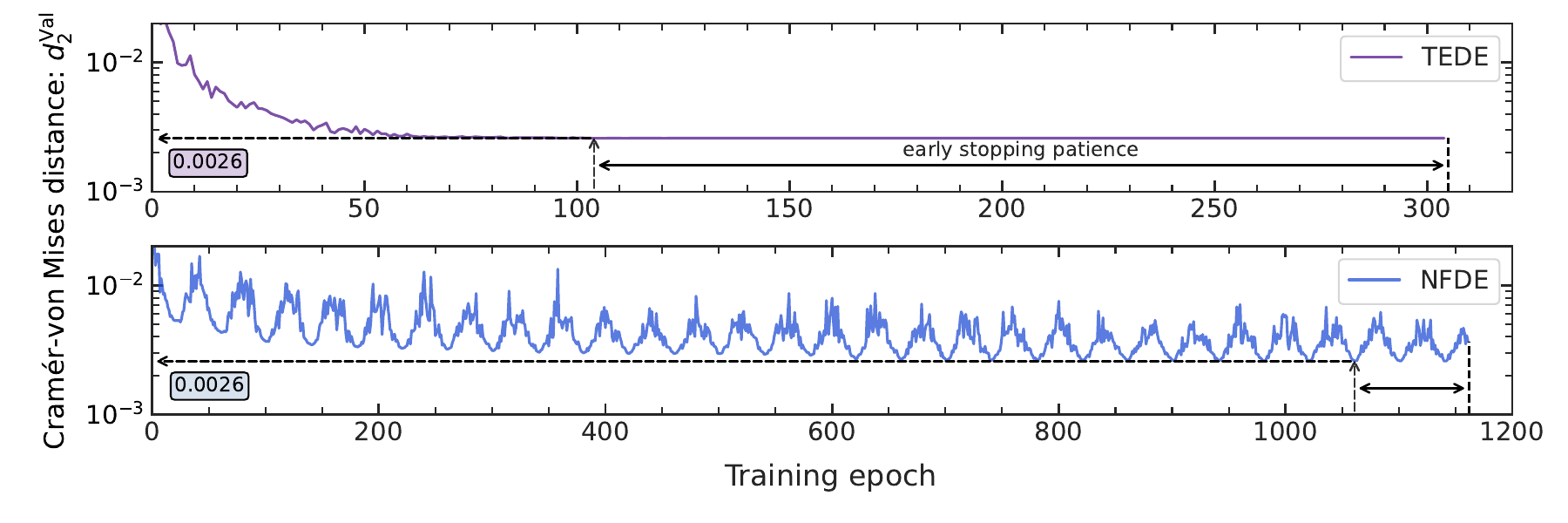}
    \caption{\textbf{Comparison of training dynamics for TEDE and NFDE models.} The plot displays the evolution of the validation Cramér-von Mises distance ($d_2^{\rm Val}$), averaged across both validation datasets, as a function of training epochs for the optimized TEDE (violet curve, top panel) and NFDE (blue curve, bottom panel) models. Lower values indicate better agreement between the model predictions and the ground truth spectra. The plot illustrates the relative learning speeds and final performance levels achieved by each approach: both models converged to a similar local minima with $d_2^{\rm Val}$ of $\sim$0.0026.
    Note that, the NFDE's training curve exhibits periodic behavior due to the use of the cosine annealing learning rate scheduler. Both training procedures were terminated using an early stopping criterion based on the stabilization of the validation metric, preventing overfitting. Early stopping's patience values (in number of epochs) are equal to 200 for TEDE and 100 for NFDE.}
    \label{fig:training_process}
\end{figure*}

In the context of machine learning, the term ``hyperparameters'' refers to parameters that define a model's architecture and its learning process. These parameters are set prior to the training process and must be optimized independently from the learnable parameters of the model (such as weights in a neural network). 

\autoref{tab:hyperparameters_comparison} presents a comprehensive summary of the hyperparameter optimization spaces and results. For both models, we conduct systematic hyperparameter optimization using Optuna~\cite{optuna} with the Tree-structured Parzen Estimator (TPE)~\cite{tpe} algorithm. TPE is an efficient Bayesian optimization approach that sequentially builds probabilistic models of the hyperparameter-to-objective function relationship, focusing search efforts on promising regions of the hyperparameter space. As the objective function for the hyperparameter search, the average of the Cramér-von Mises distances on the validation datasets is used:
\begin{equation}
d_2^{\rm Val} = \frac{1}{2} \left[\frac{1}{N_{\rm Val_1}} \sum_{i=1}^{N_{\rm Val_1}} d_{2, i}^{\rm Val_1} + \frac{1}{N_{\rm Val_2}} \sum_{j=1}^{N_{\rm Val_2}} d_{2, j}^{\rm Val_2} \right],
\end{equation}
where $\rm Val_1$ is validation dataset 1 ($N_{\rm Val_1} = 1000$ different parameter grid points) and $\rm Val_2$ is validation dataset 2 ($N_{\rm Val_2} = 3$ different parameter grid points). Thus, we select the set of hyperparameters that best minimizes $d_2^{\rm Val}$.

In addition, we employ early stopping during each optimization trial, halting the training process when the model performance on validation data stops improving for a predefined number of epochs, so-called ``patience'' (set to 200 for TEDE and 50 for NFDE).

For TEDE, we perform a comprehensive search across 250 trials, optimizing hyperparameters that can be split into two categories: 1) related directly to the model's architecture and 2) related to the training procedure. The architecture hyperparameters are focused on values defining the transformer encoder structure~\cite{transformer}: the number of tokens that encode each input MC parameter $n_{\rm tokens}$, the embedding dimension $d_{\rm model}$, the number of attention heads $n_{\rm head}$, the number of encoder layers $n_{\rm layers}$, the feed-forward dimension $d_{\rm ff}$, the dropout rate $p_{\rm drop}$, the activation function type, and the softmax temperature. As for general training hyperparameters, we explore optimizer selection (\texttt{AdamW} vs. \texttt{RMSprop}), learning rate and its scheduling strategies, weight decay, and batch size. The batch size is defined as the number of different calibration source spectra processed simultaneously. Additionally, various optimizer- and scheduler-specific hyperparameters were tuned.

For NFDE, we perform 25 trials, focusing on flow-specific hyperparameters: the number of flows ($n_{\rm flows}$) and the hidden unit dimension of the conditioning neural networks ($n_{\rm units}$), that controls their capacity. To optimize computational resource costs for the hyperparameter search space, we first performed preliminary manual optimization using a trial-and-error approach to identify the most promising hyperparameters choices and fix them. Based on these preliminary experiments, planar flows consistently outperformed radial flows across different configurations, so the flow type was fixed as planar for the systematic optimization. Several other training hyperparameters (optimizer type, base learning rate, weight decay) were also fixed based on these preliminary experiments, while others, particularly scheduler-related parameters and the $\beta_1, \beta_2$ decay rates for the \texttt{AdamW} optimizer, underwent systematic optimization.

The best found configurations (highlighted in bold in the~\autoref{tab:hyperparameters_comparison}) show the following patterns: TEDE benefited from a moderate-sized transformer with no dropout ($p_{\rm drop}=0.0$) and with the softmax temperature more than twice higher than one ($T=2.03$), that smooths the output probability distributions, and an \texttt{Exponential} learning rate scheduler. While NFDE performed best with 120 planar transformations, moderate conditioning network size ($n_{\rm units}=30$), and a \texttt{CosineAnnealing} learning rate scheduler. Both models favored the \texttt{GeLU} activation function, and both converged on similar optimization strategies with \texttt{AdamW}, although with different specific hyperparameter values ($\beta_1, \beta_2$, $T_{\rm max}$) reflecting their approach-specific differences.

After the hyperparameter optimization procedure, both models were re-trained from scratch using the resulted optimized sets of the hyperparameters. The training dynamics, which illustrates the minimization of the average Cramér-von Mises distance for the validation data, is shown for both models in~\autoref{fig:training_process}. The early stopping patience was 200 and 100 epochs for TEDE and NFDE, respectively. Note that the final patience value for NFDE model was set to be twice higher than the cosine annealing period $T_{\max}$ to ensure proper convergence through multiple learning rate cycles.

\subsection{Parameter estimation methods}
\label{subsec:parameter_estimation}

We estimate the energy response parameters ($k_B$, $f_C$, $Y$) by comparing the
ML model predictions to target data (from the testing datasets) using both
Bayesian and Frequentists inferences. In the Frequentist framework, the
\texttt{iminuit} package~\cite{iminuit} is used for cost-function minimization,
while in the Bayesian framework, we utilize a nested sampling
method~\cite{buchner2023nested} from \texttt{ultranest}~\cite{ultranest}
alongside a self-coded Metropolis-Hastings algorithm~\cite{metropolis_full, metropolis} to sample the posterior distributions.

\subsubsection{Likelihood functions for statistical inference}

Let the parameter vector be denoted by $\bm{\phi} = \{k_B, f_C, Y\}$, and the total dataset by
$\mathcal{D}$. For a specific source, we define the corresponding subset of data as
$\mathcal{D}_s$ and the overall normalization constant $N_s$ as the total number of expected events
for that source. Here and throughout this section, the index $s$ denotes a specific calibration source, taken from the following list: ${}^{137}$Cs, ${}^{40}$K, ${}^{60}$Co, ${}^{241}$Am-Be, ${}^{241}$Am-${}^{13}$C.

For both models, the likelihood function used for statistical inference quantifies the probability of observing the data given the parameters, where the parameters to be optimized are $\bm{\phi}$ and $N_s$.

\paragraph{The TEDE model (binned data).} The TEDE model is designed for binned data, where the
observation in each of the $N_b$ bins is given by $m_{s,i}$ (for $i=1,\ldots,N_b$).
Expected count of events for each bin can be found as $\mu_{s,i}(\bm{\phi})=N_s\cdot \hat{p}_{i}^{\text{TEDE}}(\bm{\phi}, S)$. The likelihood based on the Poisson distribution for a source
$s$ is given by \begin{equation}
  \mathcal{L}_s(\mathcal{D}_s\,|\,\bm{\phi}) = \prod_{i=1}^{N_b} \frac{(\mu_{s,i}(\bm{\phi}))^{m_{s,i}} \,
  e^{-\mu_{s, i}(\bm{\phi})}}{(m_{s,i})!}.
\end{equation}

Taking the natural logarithm results in the log-likelihood:
\begin{equation}
\begin{aligned}
  &\ln \mathcal{L}_s(\mathcal{D}_s\,|\,\bm{\phi}) =  \\ &= \sum_{i=1}^{N_b} \left[ m_{s,i} \cdot \ln
  \mu_{s,i}(\bm{\phi}) - \mu_{s,i}(\bm{\phi}) - \ln((m_{s,i})!) \right].
  \label{eq:binned_poisson_ll}
\end{aligned}
\end{equation}

Because the constant term $\ln(m_{s,i}!)$ does not depend on $\bm{\phi}$, it can be neglected during
optimization. For the Bayesian method, this Poisson log-likelihood is used directly in constructing
the posterior distribution. In the Frequentist approach, a related quantity known as the Poisson
log-likelihood ratio is employed. Frequently, one minimizes the test statistic defined as:
\begin{equation}
\begin{aligned}
  -2\Delta \ln \mathcal{L}_s&(\mathcal{D}_s\,|\,\bm{\phi}) =
  -2\ln\mathcal{L}_s(\mathcal{D}_s\,|\,\bm{\phi}) + 2
  \ln\mathcal{L}_s(\mathcal{D}_s\,|\,\mathcal{D}_s) = \\ &=2 \sum_{i=1}^{N_b} \left[
    \mu_{s,i}(\bm{\phi}) - m_{s,i} + m_{s,i}
  \ln\frac{m_{s,i}}{\mu_{s,i}(\bm{\phi})} \right],
\end{aligned}
\end{equation}
where the convention $m_{s,i} \ln (m_{s,i}/\mu_{s,i}(\bm{\phi})) = 0$ when $m_{s,i}=0$ is adopted.
This statistic is particularly useful when comparing nested models or evaluating goodness-of-fit in
a Frequentist framework, with the only difference from (\autoref{eq:binned_poisson_ll}) being a
data-dependent constant shift that does not affect minimization results.

\paragraph{The NFDE model (unbinned data).}
For NFDE, the analysis is performed on unbinned data. Let the individual
data points for source $S$ be denoted as $\{x_{s,i}\}_{i=1}^{M_s}$, where $M_s$ is the total
number of data points. The NFDE model outputs a PDF $\hat{p}(x_{s,i}|\phi, S)$.

With these ingredients, the extended unbinned likelihood~\cite{extended_ll_cowan,
extended_ll_barlow} can be constructed as a product of a Poisson term and terms involving the
PDF:
\begin{equation}
  \mathcal{L}_s(\mathcal{D}_s\,|\,\bm{\phi}) = \frac{N_s^{M_s} \,
e^{-N_s}}{M_s!} \cdot \prod_{i=1}^{M_s} \hat{p}(x_{s,i} | \bm{\phi}, S).
\end{equation}

Thus, the corresponding extended log-likelihood becomes:
\begin{equation}
\begin{aligned}
  \label{eq:ext_unbinned_ll}
  &\ln \mathcal{L}_s(\mathcal{D}_s\,|\,\bm{\phi}) = \\ &= M_s\ln N_{s} - N_{s} - \ln(M_s!) +
  \sum_{i=1}^{M_s} \ln \,
  \hat{p}(x_{s,i} | \bm{\phi}, S),
\end{aligned}
\end{equation}
where $\ln (M_s!)$ term may be omitted during optimization since it does not depend on $\bm{\phi}$.
This formulation naturally accounts for fluctuations in the total number of events due to the
Poisson component, while the distribution shape is governed by the PDF
term. The likelihood (\autoref{eq:ext_unbinned_ll}) is used directly in Bayesian analysis and scaled
by a factor of $(-2)$ for Frequentist analysis.

\paragraph{Extension to multiple spectra.}
Since full information originate from multiple independent
sources, our framework is required to account for their combined contribution
to parameter estimation. Using the five individual likelihood functions described earlier --- one for each calibration source --- the total likelihood function across all sources is defined as follows:
\begin{equation}
  \mathcal{L}(\mathcal{D}\,|\,\bm{\phi}) = \prod_{s=1}^{5}
\mathcal{L}_s(\mathcal{D}_s\,|\,\bm{\phi}).
\end{equation}

Similarly, since the logarithm transforms products into sums, the corresponding
log-likelihood function can be expressed as:
\begin{equation} \ln \mathcal{L}(\mathcal{D}\,|\,\bm{\phi}) = \sum_{s=1}^{5} \ln
\mathcal{L}_s(\mathcal{D}_s\,|\,\bm{\phi}). \end{equation}

We assume that data from different sources are statistically independent. Consequently, in this formulation each spectrum contributes independently to the total likelihood.

\subsubsection{Inference approaches}

After constructing the appropriate likelihood functions, the next step is to infer the parameter
vector $\bm{\phi}$. In our study, two different approaches are employed:

\paragraph{Frequentist inference.} For the Frequentist perspective, the parameter estimation is
performed by maximizing the likelihood (or equivalently, minimizing the negative logarithm of the
likelihood). The minimization is carried out with the \texttt{iminuit} package, which effectively
finds the maximum likelihood estimates (MLEs) using the cost functions defined above.

Furthermore, the uncertainties in the parameter estimates are obtained using the MINOS algorithm
implemented within \texttt{iminuit}. MINOS determines parameter uncertainties by scanning the
profile of the likelihood function around the minimum, thereby accounting for potential asymmetries
in the error estimates. Unlike simple symmetric error approximations (which are valid only when the
likelihood surface is nearly quadratic), MINOS provides robust, asymmetric uncertainties, ensuring
more reliable confidence intervals even when the likelihood is non-linear with respect to specific
parameters~\cite{minos}.

\paragraph{Bayesian inference.}
In the Bayesian approach, the inferential goal is to obtain the
posterior distribution of the parameters given the observed data. Bayes' theorem states that
\begin{equation}
  q(\bm{\phi}\,|\,\mathcal{D}) \propto \mathcal{L}(\mathcal{D}\,|\,\bm{\phi})\pi(\bm{\phi}),
\end{equation}
where $\pi(\bm{\phi})$ is the prior distribution encoding any previous knowledge about $\bm{\phi}$, which
is considered flat in our study. The specific likelihood functions used are the ones outlined above.

To sample the posterior distribution, we utilize \texttt{ultranest}, a robust nested sampling algorithm, which effectively explores the parameter space even in complex high-dimensional settings. Nested sampling works by iteratively sampling from constrained prior volumes with progressively higher likelihood thresholds, systematically shrinking the sampling region while maintaining uniform sampling within each likelihood contour. This approach enables efficient computation of both the posterior distribution and the Bayesian evidence (marginal likelihood) simultaneously. Moreover, a self-coded implementation of the Metropolis-Hastings algorithm is also applied to generate independent Markov chains, thereby providing complementary checks on the convergence and the robustness of the Bayesian inference.

The best-fit value of the parameters is determined as the median of the obtained posterior distribution, while the associated $1\sigma$ uncertainty is quantified using the $34.1\%$ of the probability on either side of the median. This ensures a statistically robust characterization of the parameter estimates, avoiding biases that could arise from asymmetric posterior distributions.

\subsubsection{Parameter estimation discussion}
The dual approach, Frequentist optimization with \texttt{iminuit} (using the MINOS algorithm for
robust uncertainty estimation) and Bayesian sampling with \texttt{ultranest} and
Metropolis-Hastings, allows for cross-validation of parameter estimates while providing deep insights
into the underlying uncertainties. The formulation of likelihood functions is inherently tied to the
data structure: binned Poisson likelihoods for the TEDE model and extended unbinned likelihoods for
the NFDE model ensure that the statistical features of the data are properly accounted for.
Together, these methods yield not only point estimates from the MLE approach but also full posterior
distributions via Bayesian inference, thus enhancing the overall robustness of our statistical
conclusions.

\section{Data Availability}
This study is based on simulation data produced by the JUNO collaboration and is not publicly available according to the collaboration policy but can be shared upon a reasonable request to the corresponding authors and with permission obtained from the collaboration.

\section{Code Availability}
The code used for this study is available at \url{https://github.com/ArseniiGav/neuromct}.

\section*{Acknowledgments}
We are thankful to the JUNO collaboration for the support and advice provided during the drafting of this manuscript. We are also very grateful to CNAF and JINR cloud services for providing the computing resources necessary for the simulated data production and to CloudVeneto for offering IT support and infrastructure for training the machine learning models used in this study. Arsenii Gavrikov has received funding from the European Union's Horizon 2020 research and innovation programme under the Marie Skłodowska-Curie Grant Agreement No. 101034319 and from the European Union --- NextGenerationEU. Dmitry Dolzhikov and Maxim Gonchar are supported in the framework of the State project ``Science'' by the Ministry of Science and Higher Education of the Russian Federation under the contract 075-15-2024-541.

\bibliography{references}

\onecolumngrid
\clearpage
\appendix

\appendix

\section{Energy response parameters effects with the TEDE model}
\label{ax:tede_param_dependencies}

\begin{figure*}[!htb]
	\centering
	\includegraphics[width=1\textwidth]{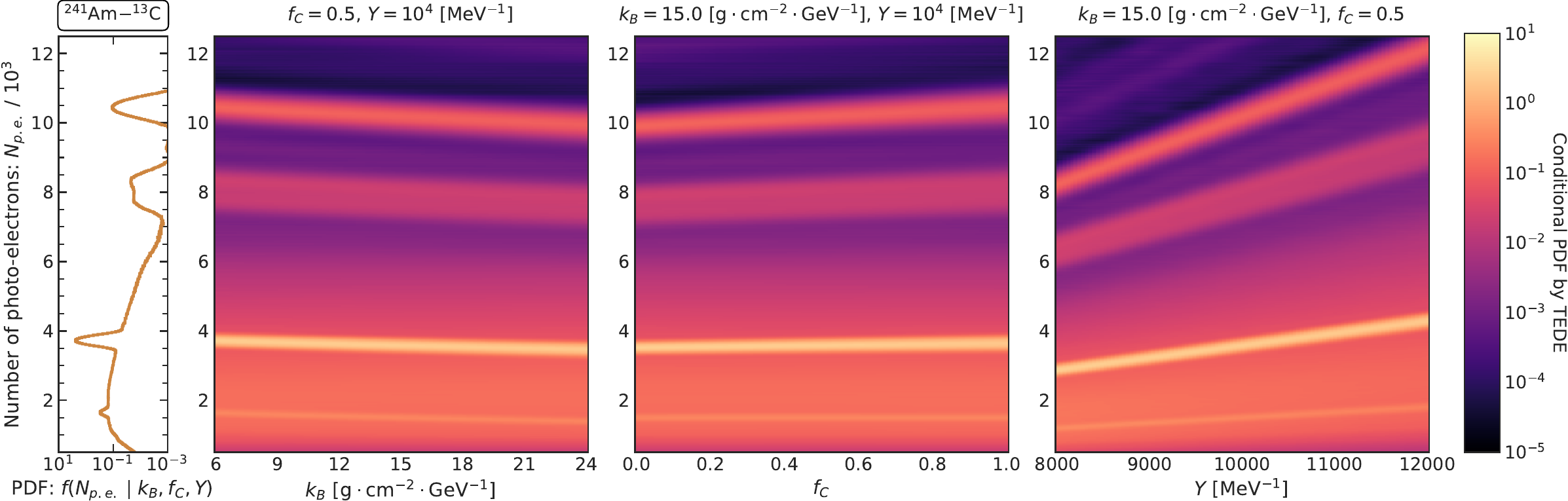}
	\caption{\textbf{Visualization of how the energy response parameters impact the calibration source energy spectra, shown through conditional PDF modeled by TEDE.} Heatmaps present the probability density given by the TEDE model for the ${}^{241}$Am-${}^{13}$C neutron source as each parameter is varied independently: $k_B$ (left), $f_C$ (middle), and $Y$ (right). The rotated spectrum on the left shows the modeled by TEDE spectrum of ${}^{241}$Am-${}^{13}$C for illustration; the following parameter values were used: $k_B = 6.0 \text{ g}\cdot \text{cm}^{-2} \cdot \text{GeV}^{-1}$, $f_C = 0.5$, $Y = 10000 \text{ MeV}^{-1}$. For each panel, the two parameters that are not varied are fixed at their median grid values. Color intensity represents conditional PDF values on a logarithmic scale, showing the expected parameter-specific effects. Note, that the y-axis upper limit of 12.5k $N_{p.e.}$ is chosen for visualization purposes only, as it encompasses the main parts of the spectra. Unlike NFDE's continuous results presented in~\autoref{fig:conditional_pdf_heatmap_nfde_AmC}, the conditional PDFs modeled by TEDE are discrete due to the histogram-based approximation.}
    \label{fig:conditional_pdf_heatmap_tede_AmC}
\end{figure*}

\begin{figure*}[!htb]
	\centering
	\includegraphics[width=1\textwidth]{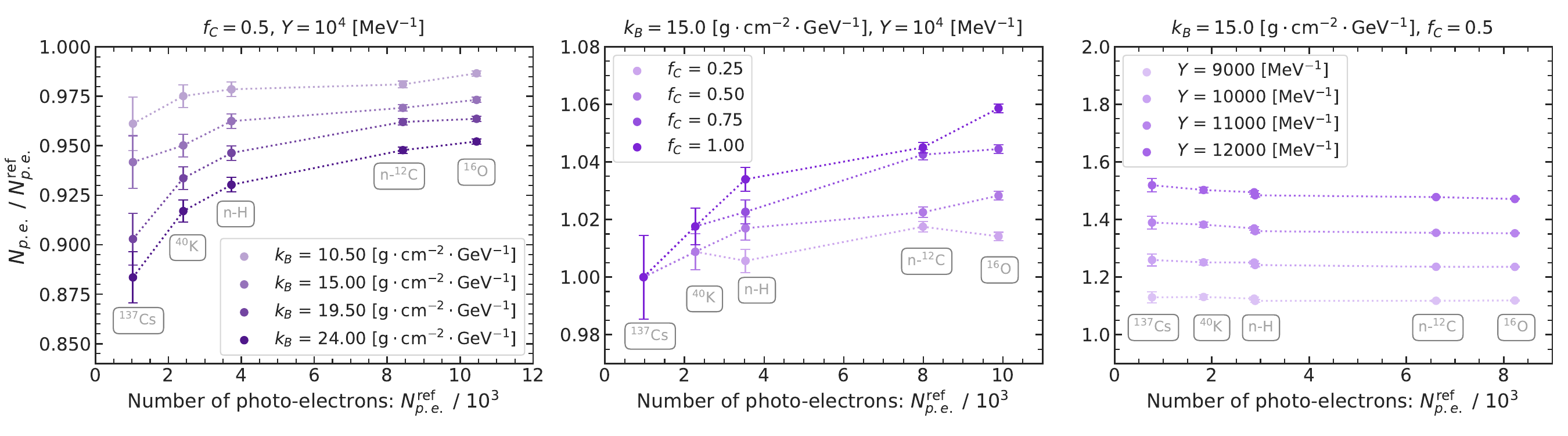}
    \caption{\textbf{The effect of the parameters as a function of energy.} This plot demonstrates how each parameter affects the relative positions of the main peaks across the energy scale of the calibration sources. The main peaks of the spectra from multiple calibration sources serve as standard candles: $^{137}$Cs, $^{40}$K, and neutron capture peaks (n-H, n-$^{12}$C) as well as the $^{16}$O excitation. Moving along the y-axis shows the same gamma peak under different energy response scenarios (different parameter values), while moving along the x-axis shows gamma peaks (i.e., different energies) under the same energy response scenario. For each parameter ($k_B$, $f_C$, and $Y$), five values spanning the full parameter range were selected while keeping the other two parameters fixed at the mean values of the grid. The y-axis shows the relative shift in peak positions compared to references. The plot also highlights the different influence of each parameter: $k_B$ shows increasing non-linearity at lower energies, $f_C$ exhibits a greater effect at higher energies, while $Y$ demonstrates a nearly constant scaling with slight deviations at low energies due to dark noise contributions. Unlike NFDE's continuous results presented in~\autoref{fig:nl_nfde}, the parameter effect estimations by TEDE are limited by the histogram discretizations.}
    \label{fig:nl_tede}
\end{figure*}

\clearpage
\section{Parameter estimation results with the MIGRAD+MINOS method}
\label{ax:param_estimation_other_methods}

\begin{figure*}[!htb]
  \centering
  \includegraphics[width=1\textwidth]{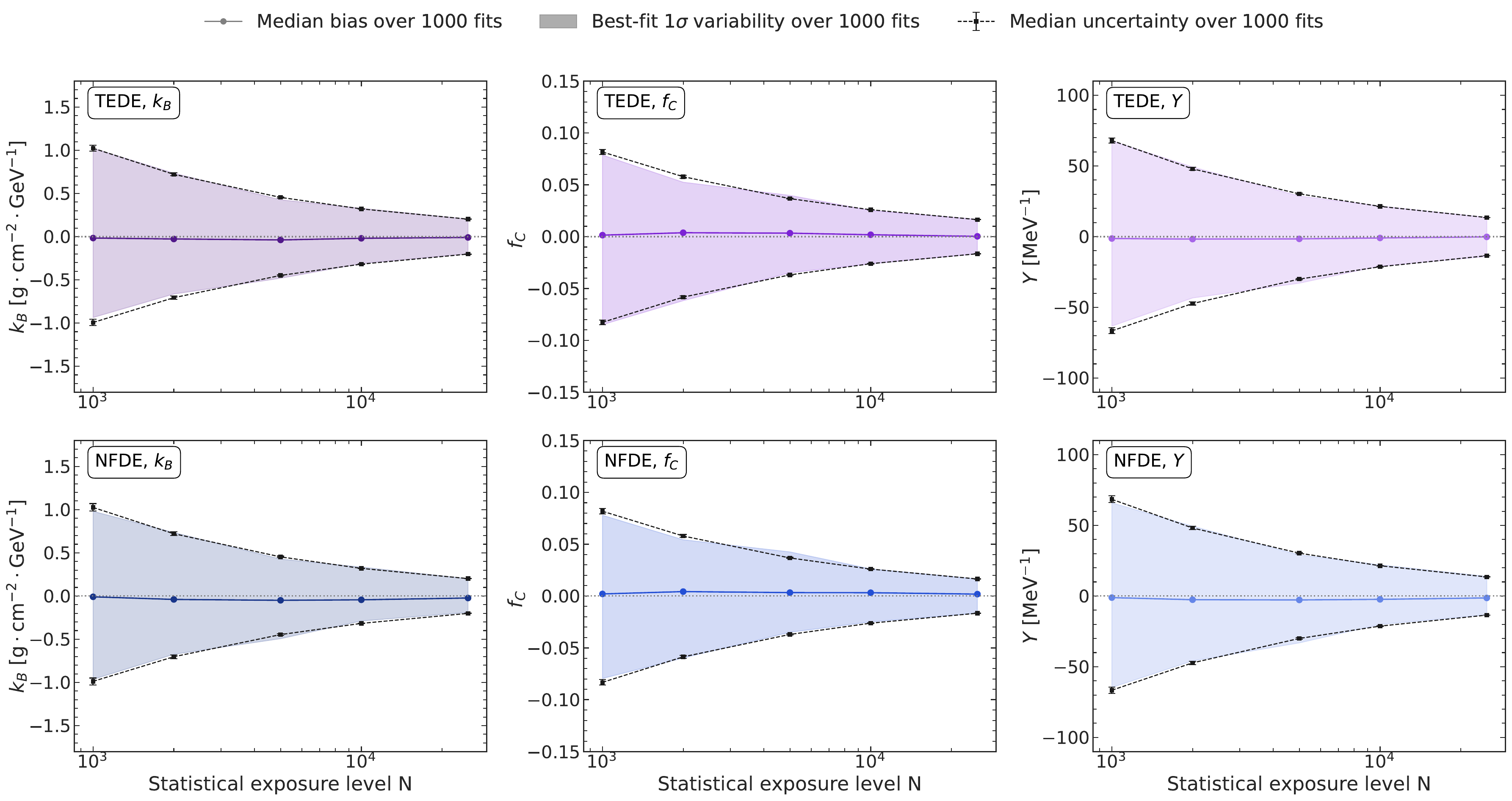}
  \includegraphics[width=1\textwidth]{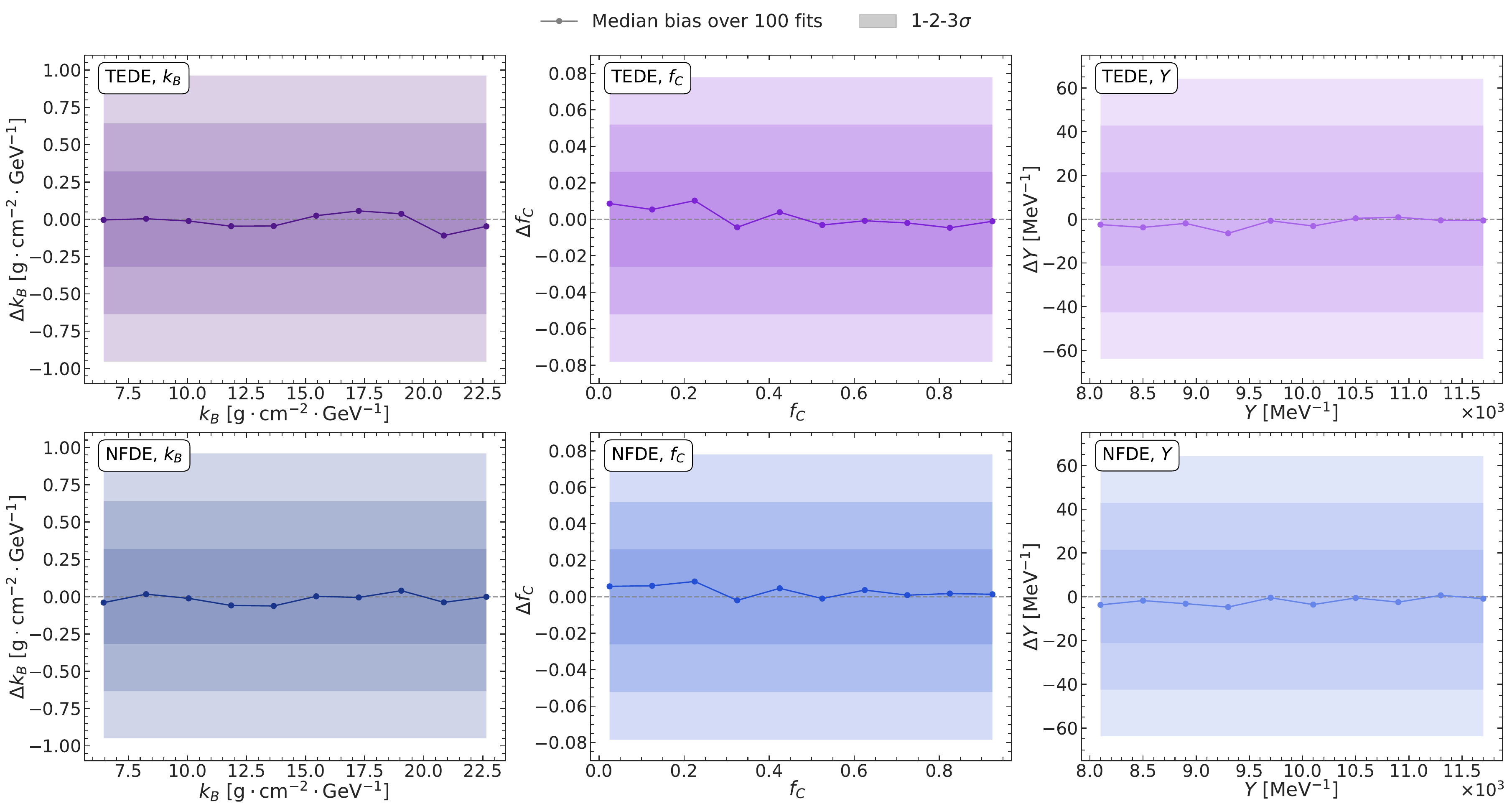}
\caption{
  \textbf{Results obtained using the MIGRAD+MINOS method.} This plot presents results obtained using the MIGRAD+MINOS algorithms. The notation follows the same conventions as those used for the Bayesian nested sampling (\texttt{ultranest}) results shown in Figures \ref{fig:results_td2} and \ref{fig:results_td1}. These findings are consistent with the results obtained using both Bayesian nested sampling (\autoref{fig:results_td2} and \autoref{fig:results_td1}) and the Metropolis-Hastings MCMC method (\autoref{fig:results_mh}), demonstrating agreement among all three parameter estimation methods.
}
\label{fig:results_im}
\end{figure*}

\clearpage
\section{Parameter estimation results with the Metropolis-Hastings MCMC method}
\begin{figure*}[!htb]
  \centering
  \includegraphics[width=1\textwidth]{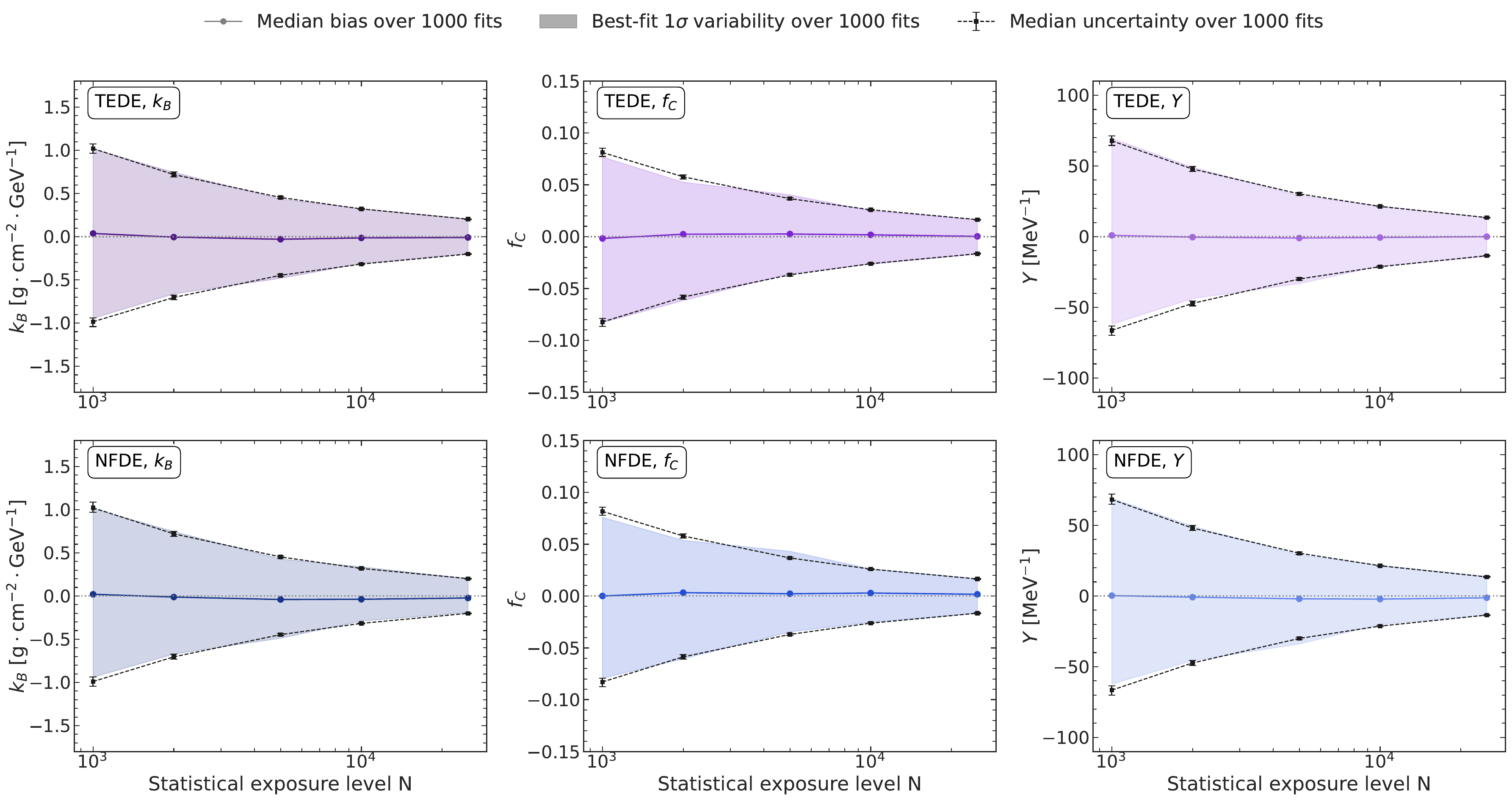}
  \includegraphics[width=1\textwidth]{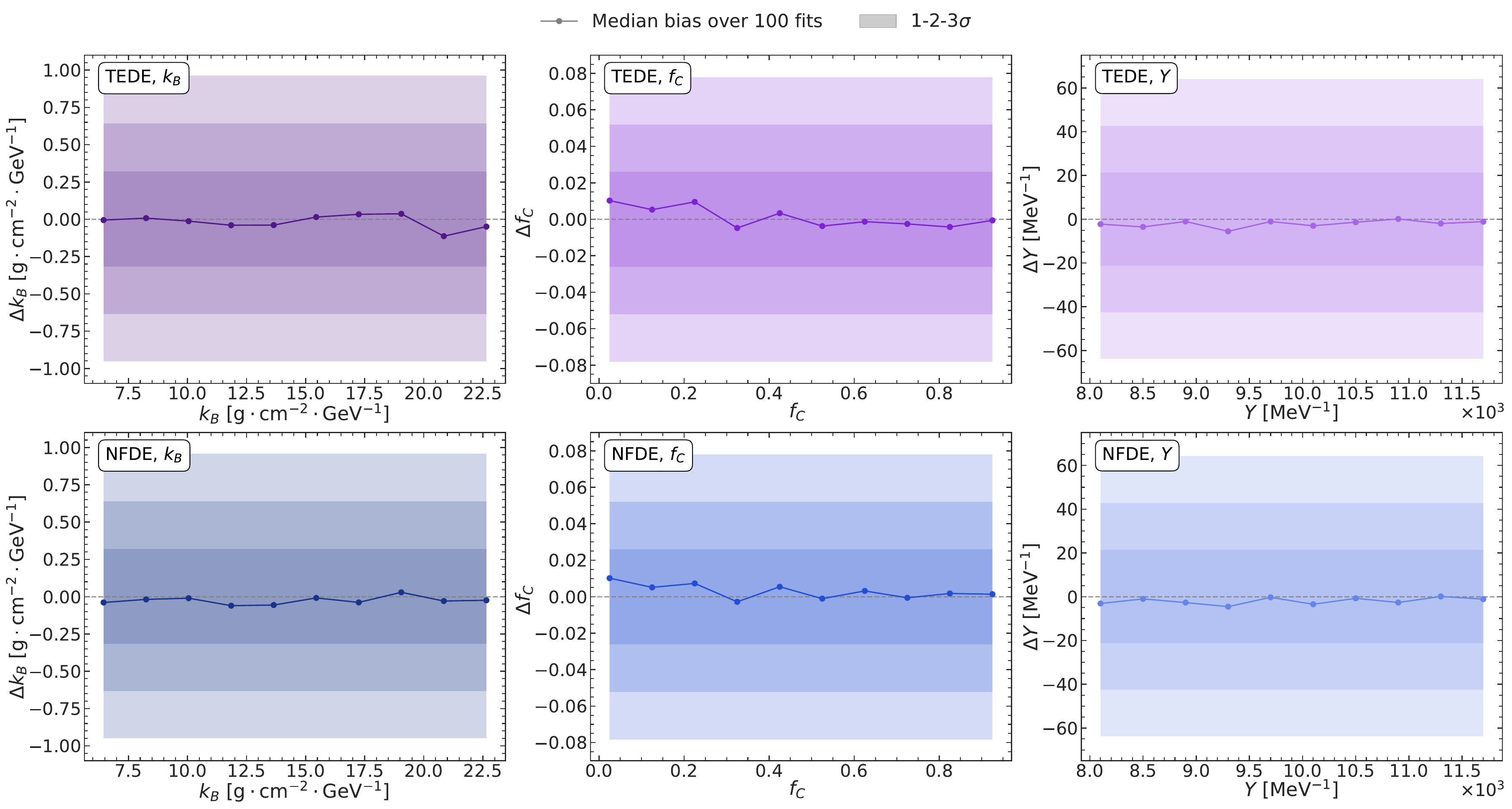}
\caption{
  \textbf{Results obtained using the Metropolis-Hastings MCMC method.} This plot presents results obtained using the Metropolis-Hastings MCMC algorithm. The notation follows the same conventions as those used for the Bayesian nested sampling (\texttt{ultranest}) results shown in Figures \ref{fig:results_td2} and \ref{fig:results_td1}. These findings are consistent with the results obtained using both Bayesian nested sampling (Figures \ref{fig:results_td2} and \ref{fig:results_td1}) and MIGRAD+MINOS method (Figure \ref{fig:results_im}), demonstrating agreement among all three parameter estimation methods.
}
\label{fig:results_mh}
\end{figure*}

\end{document}